%% file: subtraction-nlo.tex
\documentclass[twocolumn, final, epjc3]{svjour3}  
\pdfoutput=1

\smartqed
\journalname{Eur. Phys. J. C}

\RequirePackage[utf8]{inputenc}
\RequirePackage{xspace}
\RequirePackage{graphicx}
\RequirePackage{amsmath}
\RequirePackage{amssymb}
\RequirePackage[unicode]{hyperref}
\RequirePackage[margin=2cm]{geometry}
\RequirePackage{xcolor}
\RequirePackage{hepunits}
\RequirePackage{enumitem}
\RequirePackage{cuted}
\RequirePackage{etoolbox}
\RequirePackage{esint}
\usepackage{tikz}
\usepackage[compat=1.1.0]{tikz-feynman}

\RequirePackage{definitions} 

\begin{document}

\title{Proton internal pressure from deeply virtual Compton scattering on collider kinematics}

\author{
H.~Dutrieux\thanksref{email1,address1}
\and
T.~Meisgny\thanksref{address2}
\and
C.~Mezrag\thanksref{email3, address2} 
\and
H.~Moutarde\thanksref{email4,address2}
}

\thankstext{email1}{e-mail: hldutrieux@wm.edu}
\thankstext{email3}{e-mail: cedric.mezrag@cea.fr}
\thankstext{email4}{e-mail: herve.moutarde@cea.fr}


\institute{%
  Physics Department, William \& Mary, Williamsburg, VA 23187, USA \label{address1}
  \and
Irfu, CEA, Universit\'e Paris-Saclay, F-91191 Gif-sur-Yvette, France \label{address2}
}

\date{Received: date / Accepted: date}

\maketitle

\sloppy

\begin{abstract}
The unique experimental connection to the QCD energy-momentum tensor offered by generalised parton distributions has been strongly highlighted in the past few years  with attempts to extract the pressure and shear forces distributions within the nucleon.
If, in principle, this can be performed in a model independent way from experimental data, in practice,
the current limited precision and kinematic coverage make such an extraction very challenging.
Moreover, the limitation to a leading-order description in the strong coupling of the data has provided only an indirect and weakly sensitive access to gluon degrees of freedoms, solely through their mixing to quarks via evolution.
In this paper we address this issue by providing a next-to-leading order formalism allowing a reanalysis of global fits with genuine gluonic degrees of freedom.
In addition, we provide an estimate of the reduction in uncertainty that could stem from the extended kinematic range relevant for the future Electron Ion Collider. Finally, we stress the connection between the analysis of the dispersion relation in terms of generalised parton distributions and the deconvolution problem.
\end{abstract}

\keywords{3D Nucleon Structure \and Nucleon Tomography \and Global Fit \and Deeply Virtual Compton Scattering \and DVCS \and Compton Form Factor \and CFF \and Next-to-leading order \and NLO \and Dispersion Relation \and Subtraction Constant \and Generalised Parton Distribution \and GPD \and Deconvolution problem \and Artificial Neural Network \and EIC \and EicC \and Jefferson Lab \and PARTONS Framework}
\PACS{12.38.-t \and 13.60.-r \and 13.60.Fz \and 14.20.-c}


\section{Introduction}
\label{sec:introduction}

\input{sec_introduction.tex}

\section{EMT experimental access through GPDs}
\label{sec:EMT}

\input{sec_EMT.tex}

\section{DVCS dispersion relations beyond leading order}
\label{sec:dispersion-relations}

\input{sec_DR.tex}

\section{Extraction of the pressure distribution on collider kinematics: an inverse problem}
\label{sec:pressure-distributions-nlo}

\input{section4.tex}

\section{Results on current experimental data}
\label{sec:pressure-shadow} 

\input{section5.tex}

\section{An EIC perspective}

\input{section6.tex}

\section{Conclusion}
\label{sec:conclusion}

We have provided a re-derivation of dispersion relations for DVCS at all order in perturbation theory. Our presentation highlights that dispersion relations contain far more information than the commonly acknowledged restriction to the $D$-term. In fact, eq. \eqref{eq:SymDispersionRelation} along with eq. \eqref{eq:Hexpansion} give a very synthetic picture of the information that can possibly be extracted from an arbitrary knowledge of CFFs at a given scale. We have then stressed that the scale dependence of measurements is crucial to mitigate the issues related to the deconvolution problem. An intuitive presentation in terms of \textit{shadow} $D$-terms allows to construct simple estimates of the impact on the uncertainty of the range in scales on which DVCS is measured.

Using those tools, we have re-considered the dispersion relation global analysis of Ref. \cite{Dutrieux:2021nlz}. We have proposed a new statistical method aiming at giving a sounder account of correlated fits in the presence of outliers. Our re-analysis of the LO $D$-term gives however very similar results to the previous publication. We extend the analysis to NLO. We find that the NLO results are generally fairly similar to the LO. The major modification comes when an explicit gluon $D$-term is allowed to be freely fitted on the data. Then we find at NLO a very large correlation between the fitted quarks and gluons that does not exist at LO. We provide an explanation for this fact, linked to the similarity of scale dependence of the evolution operators.

Using the simple formalism of shadow $D$-terms, we finally establish generic estimates of the reduction of uncertainty that one could expect from the range of scales probed at an EIC. We find that extending the region of measurements with a similar statistical accuracy as current best measurements to 20 GeV$^2$ could bring a reduction of uncertainty in the deconvolution problem by a factor 2 to 5 depending on the quantity of interest. One should however keep in mind that the EIC will additionally reduce the statistical uncertainty on the subtraction constant, by measuring CFFs at values of Bjorken-$x$ where they are poorly constrained so far. Therefore, the potential EIC impact on our experimental knowledge of $d_1$ could be larger and remains to be fully estimated. In the mean-time, we expect the knowledge of the pressure and shear forces within the nucleon to be mostly driven by lattice-QCD calculations.


\begin{acknowledgements}
The authors thank V. Bertone, C. Lorcé, K. Orginos, J. Rodriguez-Quintero, E. Romero Alcalde and P. Sznajder for valuable discussions. This project was supported by the European Union's Horizon 2020 research and innovation programme under grant agreement No 824093, and with the support of the French Government scholarships programme. This research was funded, in part, by l’Agence Nationale de la Recherche (ANR), project ANR-23-CE31-0019, supported by the GLUODYNAMICS project funded by “P2IO LabEx (ANR-10-LABX-0038)” in the framework of Investissements d’Avenir (ANR-11-IDEX-0003-01), managed by the Agence Nationale de la Recherche (ANR), France. HD was supported by U.S. DOE Grant \#DE-FG02-04ER41302 and under the Laboratory Directed Research and Development Program (LDRD 2412) at the Thomas Jefferson National Accelerator Facility for the U.S. Department of Energy. For the purpose of open access, the author has applied a CC-BY public copyright licence to any Author Accepted Manuscript (AAM) version arising from this submission. 
\end{acknowledgements}

\appendix

\section{Open source resources}
\label{appendix:opensource_code}

The neural network global fit of CFFs used in this analysis is available in the PARTONS framework \cite{Berthou:2015oaw}. 
The code of this framework is open source and can be found online at \url{https://drf-gitlab.cea.fr/partons/core/partons} on version 3 of the GPL (GPLv3).

\bibliographystyle{unsrt}
\bibliography{Bibliography}

\end{document}

%% file: sec_introduction.tex
More than a century after its discovery by E. Rutherford \cite{Rutherford:1919fnt}, the proton is still at the core of an intense research activity.
Among other aspects, its mechanical properties, described through the quantum-chromodynamics (QCD) energy-momentum tensor (EMT), have attracted a significant attention in the last decade yielding many theoretical (see for instance \cite{Polyakov:2018zvc,Lorce:2018egm,Freese:2021jqs}), phenomenological (for instance \cite{Burkert:2018bqq,Kumericki:2019ddg,Dutrieux:2021nlz}), lattice (as in \cite{Alexandrou:2017oeh,Alexandrou:2020sml,Wang:2021vqy}) and continuum studies (see \emph{e.g.} \cite{Nair:2024fit,Yao:2024ixu}).
The reason for this interest is that the macroscopic properties of the proton such as its mass or its spin are expected to be emergent phenomena from the microscopic interaction between quarks and gluons.
The computation of the quark and gluon contributions to the macroscopic properties of the proton, and the comparison with the experimental extraction of these contributions has become one of the main objectives of modern hadron physics.

The connection to experimental data is certainly one of the main factors explaining the recent interest for the EMT.
Indeed, it was shown almost three decades ago \cite{Ji:1996ek} that one can build an indirect experimental access to the EMT via generalised parton distributions (GPDs) \cite{Mueller:1998fv,Ji:1996ek,Ji:1996nm,Radyushkin:1996ru,Radyushkin:1997ki}.
The latter enter the description of deep exclusive processes according to QCD factorisation theorems \cite{Collins:1996fb,Collins:1998be}.
One can for instance highlight deeply virtual Compton scattering (DVCS) \cite{Ji:1996nm}, timelike Compton scattering (TCS) \cite{Berger:2001xd}, deep virtual meson production (DVMP) \cite{Mueller:2013caa}, multiparticle production \cite{Boussarie:2016qop,Duplancic:2018bum,Grocholski:2021man} or single diffractive hard exclusive processes \cite{Qiu:2022bpq,Qiu:2022pla,Qiu:2023mrm}.
However, the main source of experimental information remains today DVCS which was measured in several facilities in the last two decades (see \emph{e.g.} \cite{Girod:2007aa,HERMES:2012gbh,Jo:2015ema,Defurne:2015kxq,Defurne:2017paw,COMPASS:2018pup,JeffersonLabHallA:2022pnx,CLAS:2022syx} and which is currently the core of an intense experimental program at the Thomas Jefferson Laboratory in the USA and at COMPASS at CERN.

This experimental effort has triggered an important theoretical interest for DVCS, which is today the deep exclusive process with the clearest theoretical framework.
Higher order corrections up to next-to-next to leading order (NNLO) have been derived \cite{Braun:2020yib}.
Higher-twist kinematic corrections are also available \cite{Braun:2012hq}.
However, despite its mature theoretical description, DVCS does not allow to extract unambiguously GPDs.
The reason is to be found in the so-called deconvolution problem \cite{Bertone:2021yyz,Dutrieux:2021wll}, that is the ill-definedness of the inverse problem relating DVCS form factors to GPDs, embodied by the notion of shadow GPDs.
One form factor of the EMT has been particularly studied since it can be accessed from a dispersive analysis of DVCS \cite{Anikin:2007yh} without requiring the full extraction of the GPDs. Indeed, this form factor only depends on the Polyakov-Weiss $D$-term \cite{Polyakov:1999gs}, accessible directly from the real and imaginary parts of the Compton Form Factors. However, this extraction is also plagued by shadow $D$-term contributions, hinted at in \cite{Dutrieux:2021nlz} and presented in greater detail here.

In this paper, we assess the feasibility of extracting from DVCS data independently genuine quarks and gluon contributions to the pressure and shear forces inside the proton, taking into account shadow $D$-terms and evolution.
After introducing our notations and conventions in section 2, we propose in section 3 a new derivation of the dispersion relations of DVCS amplitudes at any order of perturbation theory.
This presentation highlights that dispersion relations can provide more information than solely the canonical one concerning the $D$-term. We present the issue of the deconvolution problem and introduce the notion of shadow $D$-terms in section 4. Then, we apply our formalism on an existing global fit in section 5 and present the results of the first next-to-leading order extraction.
In section 6, we investigate more closely the impact of shadow $D$-term on a kinematic range relevant at future facilities, in view of the future electron-ion collider (EIC).


%% file: sec_EMT.tex
The proton matrix element of the local gauge-invariant QCD energy-momentum tensor (EMT) operator can be parameterised in terms of five gravitational form factors (GFFs) $A_{a}(t)$, $B_{a}(t)$, $C_{a}(t)$, $\bar C_{a}(t)$ and $D_{a}(t)$ \cite{Bakker:2004ib,Leader:2013jra} where $t=\Delta^2$ with $\Delta=p'-p$ the four-momentum transfer to the proton (see Ref. \cite{Dutrieux:2021nlz}). The label $a$ denotes either the quark flavour ($a=q$) or the gluon ($a=g$) contribution to the EMT. These GFFs allow to define various distributions of so-called mechanical properties of the proton, like distributions of pressure and shear stress induced by the nucleon's partonic structure \cite{Polyakov:2002yz,Polyakov:2018zvc,Lorce:2018egm}. Some of these GFFs are accessible thanks to their remarkable relation to generalised parton distributions (GPDs) introduced in \cite{Mueller:1998fv,Ji:1996ek,Ji:1996nm,Radyushkin:1996ru,Radyushkin:1997ki}, such as \cite{Diehl:2003ny}:
\begin{align}
  \label{eq:mel1H}
  \int_{-1}^{1}\ud x\,x^{1-p_a}\,H^a(x,\xi,t)&=A_a(t)+4\xi^2C_a(t) \,, \\
  \label{eq:mel1E}
  \int_{-1}^{1}\ud x\,x^{1-p_a}\,E^a(x,\xi,t)&=B_a(t)-4\xi^2C_a(t) \,,
\end{align}
where $p_q = 0$, $p_g = 1$ and $H^a(x, \xi, t)$ and $E^a(x, \xi, t)$ are leading-twist chiral-even GPDs depending on $x$, the average longitudinal light-front momentum fraction of the active parton and $\xi$, the skewness variable describing the transfer of longitudinal light-front momentum to the system. The link between GFFs and GPDs offers a unique opportunity for an experimental access to the mechanical properties of hadron matter, thanks to the sensitivity to GPDs of a wide class of exclusive experimental channels.
We also highlight that the specific polynomial $\xi$ dependence of the Mellin moments of eqs. \eqref{eq:mel1H} and \eqref{eq:mel1E} is in fact a general property called polynomiality \cite{Ji:1998pc,Radyushkin:1998bz} and is generalised to higher moments as:
\begin{align}
  \label{eq:polynomialityH}
  \int_{-1}^1\ud x\,x^{m-p_a} H^a(x,\xi,t) & = \sum_{j=0}^{\left[\frac{m}{2}\right]} A_{m;2j}^a(t) (2\xi)^{2j}\nonumber \\
  &\hspace{-20pt}\quad  +\textrm{mod}(2,m)(2\xi)^{m+1}C_m^a(t),\\
  \label{eq:polynomialityE}
  \int_{-1}^1\ud x\,x^{m-p_a} E^a(x,\xi,t) & = \sum_{j=0}^{\left[\frac{m}{2}\right]} B_{m;2j}^a(t) (2\xi)^{2j}\nonumber \\
&\hspace{-20pt}\quad  -\textrm{mod}(2,m)(2\xi)^{m+1}C_m^a(t),
\end{align}
where $[\dots]$ is the floor function and $\textrm{mod}(2,m)$ is 0 for $m$ even and 1 for $m$ odd. The polynomiality property is equivalent to the so-callled double distribution formalism introduced independently in \cite{Mueller:1998fv,Radyushkin:1997ki} (see also \cite{Chouika:2017dhe,Mezrag:2022pqk} for a modern picture of the connection between the two). The double distributions $F^a$ and $K^a$ are connected to the GPDs through:
\begin{align}
  \label{eq:DDF}
  H^a(x,\xi,t) = & \int_\Omega \textrm{d}\beta \textrm{d}\alpha \bigg[\beta^{p_a}F^a(\beta,\alpha,t) \nonumber \\
    &\hspace{-20pt} +\xi^{1+p_a} D^a(\alpha,t)\delta(\beta) \bigg] \times \delta(x-\beta-\alpha\xi),\\
    \label{eq:DDK}
  E^a(x,\xi,t) = & \int_\Omega \textrm{d}\beta \textrm{d}\alpha \bigg[\beta^{p_a} K^a(\beta,\alpha,t)\nonumber \\
&\hspace{-20pt}-\xi^{1+p_a} D^a(\alpha,t)\delta(\beta) \bigg]  \times \delta(x-\beta-\alpha\xi),
\end{align}
with $\Omega = \{(\alpha,\beta)| |\alpha|+|\beta|\le 1\}$ and where $D^a$ is the so-called Polyakov-Weiss $D$-term whose first Mellin moments yields the GFF $C_a(t)$ in eq. \eqref{eq:mel1H}:
\begin{align}
    \label{eq:eihwjcnalkxm}
  C_a(t) = \frac{1}{4}\int_{-1}^{1}\ud \alpha\,\alpha^{1-p_a} D^a(\alpha,t)\,.
\end{align}

However, the unambiguous model-independent extraction of GPDs from one of the most promising current channels, namely deeply virtual Compton scattering (DVCS), has already been demonstrated to be practically unfeasible in \refcite{Bertone:2021yyz}.
The reason is to be found in the relation between the DVCS amplitude, parametrised with Compton Form factors ($\mathcal{H}$, $\mathcal{E}$, \dots ) and GPDs:
\begin{align}
  \label{eq:CFFquarkDef}
  \mathcal{H}^a(\xi,t,Q^2) &= \int_{-1}^1 \frac{\textrm{d}x}{\xi} T^a\left(\frac{x}{\xi},\frac{Q^2}{\mu^2},\alpha_s \right) \frac{H^a(x,\xi,t,\mu^2)}{\xi^{p_a}},
\end{align}
where $T^a$ is the DVCS coefficient function. The convolution reveals itself not to be numerically invertible on any relevant range in $Q^2$, the virtuality of the photon mediating the interaction between the lepton beam and the proton target in DVCS, yielding out of control uncertainties. This situation can partly be tamed by exploiting the theoretical constraints applied on GPDs \cite{Dutrieux:2021wll} (see also \cite{Chavez:2021llq} for a exhaustive list of these properties), but theoretical uncertainties remain significant.

In this context, the GFF $C_a(t)$ has attracted a specific interest since it does not require a full extraction of GPDs, but is instead sensitive to the $D$-term only. The latter can be probed more specifically in a dispersive formalism of DVCS \cite{Teryaev:2005uj,Anikin:2007yh,Diehl:2007jb}. A careful analysis of the world DVCS data using this dispersive formalism at leading order (LO) was performed in \refcite{Dutrieux:2021nlz} using realistic uncertainties on DVCS form factors coming from a neural network analysis \cite{Moutarde:2019tqa}. The dependence on $Q^2$ was taken into account through the use of evolution equations for the scale dependence of the $D$-term, but the LO approach did not take into account any direct gluon contribution to the subtraction constant. It is our objective in this paper to propose a full next-to-leading order (NLO) treatment, whose relevance becomes stringent with the lever arm in $Q^2$ promised by future collider experiments, in particular those to be conducted in the electron-ion collider (EIC) \cite{Accardi:2012qut,AbdulKhalek:2021gbh}, Chinese electron-ion collider (EicC) \cite{Chen:2018wyz,Anderle:2021wcy} or large hadron-electron collider (LHeC) \cite{LHeCStudyGroup:2012zhm}.


%% file: sec_DR.tex
In this section, we provide a brief summary of the state-of-the-art regarding DVCS dispersion relations and provide an alternative proof beyond leading order. We also generalise the dispersion relation for an arbitrary number of subtractions, allowing a new way to connect moments of GPDs with experimental data.
Note however that we focus only on $s$ and $u$ channel dispersion relations, constraining the $x$ and $\xi$ dependence of GPDs. $t$-channel dispersion relations have been investigated in \cite{Pasquini:2014vua}, and are beyond the scope of this paper.

\subsection{State of the art and dispersion relations at LO and NLO}

Dispersion relations at Born order were first derived in ref. \cite{Teryaev:2005uj,Anikin:2007yh}. The authors took advantage of the explicit leading-order expression of the CFF in terms of GPDs to derive the dispersion relations. In a nutshell, at LO, the connection between GPDs and CFFs are given by:
\begin{align}
  \label{eq:LOReCFF}
  \Re \mathcal{H}^q(\xi) & \eqLO e_q^2\fint_{-1}^1H^q(x,\xi)\left[\frac{1}{\xi-x}-\frac{1}{\xi+x}\right]\,\ud x ,\\
   \label{eq:LOImCFF}
  \Im \mathcal{H}^q(\xi) & \eqLO \pi e_q^2 \left[H^q(\xi, \xi)-H^q(-\xi, \xi)\right] ,
\end{align}
where $\fint$ indicates that the integrals are regularised through the Cauchy principal value prescription.  Using the dispersion relation relating the real and imaginary part of the Compton Form Factor:
\begin{align}
  \label{eq:SubtractionLO}
  \Re \mathcal{H}^q(\xi) = \frac{1}{\pi} \fint_{0}^1 \ud x \Im \mathcal{H}^q(x) \left[\frac{1}{\xi-x}-\frac{1}{\xi+x}\right] + \mathcal{S}^q(\xi)
\end{align}
one introduces $\mathcal{S}^q$, the so-called subtraction constant associated with a flavour $q$.
Combining eqs. \eqref{eq:LOReCFF}, \eqref{eq:LOImCFF} and \eqref{eq:SubtractionLO}, one gets the following expression for the subtraction constant at LO:
\begin{align}
  \mathcal{S}^q(\xi) & \eqLO  e^2_q\fint_{-1}^1\frac{H^q(x,\xi)-H^q(x,x)}{\xi-x}\,\ud x \nonumber \\
&\hspace{30pt} -e^2_q\fint_{-1}^1\frac{H^q(-x,\xi)-H^q(-x,x)}{\xi-x}\,\ud x  \nonumber \\
  & \eqLO   2e^2_q \int_{0}^1 [H^q(x, \xi) - H^q(x,x)] \nonumber \\
  & \hspace{30pt}\quad \times \bigg[\frac{1}{\xi-x} - \frac{1}{\xi+x}\bigg]\, \ud x
\end{align}
Expanding $H^q(x,\xi)$ as a Taylor series around $x=\xi$, and using the polynomiality condition \eqref{eq:polynomialityH} or equivalently the DD representation \eqref{eq:DDF}, one recovers the well-known leading-order relation between the subtraction constant and the $D$-term:
\begin{align}
  \label{eq:LOsubtraction}
  \mathcal{S}^q(t) \eqLO 2e_q^2\int_{-1}^1 \textrm{d}z \frac{D^q(z,t)}{1-z}.
\end{align}

DVCS dispersion relations beyond the Born order were considered in ref. \cite{Diehl:2007jb}
whose analysis is based on the dispersion relation of eq. \eqref{eq:SubtractionLO}.
 Then, introducing GPDs through the leading-twist all-order factorisation theorem (eq. \eqref{eq:CFFquarkDef}) as well as the polynomiality property, they derive the general form of the subtraction constant as:
\begin{align}
  \label{eq:DiehlDR}
  \mathcal{S}^q = \frac{2}{\pi} \fint_1^\infty \textrm{d}\omega\, \Im T^q(\omega,\frac{Q^2}{\mu^2},\alpha_s) \int_{-1}^1 \textrm{d}\alpha \frac{D^q(\alpha,t,\mu^2)}{\omega - \alpha}.
\end{align}
To the best of our knowledge, this is the sole study on the topic of dispersion relations beyond Born order until now.

This integral generalises the expression of the subtraction constant at higher order in perturbation theory, at the price of introducing a second integration variable, to be integrated to infinity.
In the following, we present another approach to generalise dispersion relations at higher orders, allowing us to write the subtraction constant as a single integral over the $D$-term.

\subsection{Analytic properties of the Compton Form Factors}

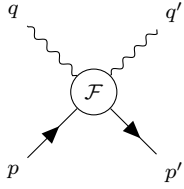
\begin{figure}[!h]
	\centering
	\begin{tikzpicture}
		\begin{feynman}
			\vertex[draw, circle] (A) {\(\mathcal{F}\)};
			\vertex[above left of=A] (i1) {\(q\)};
			\vertex[below left of=A] (i2) {\(p\)};
			\vertex[above right of=A] (o1) {\(q'\)};
			\vertex[below right of=A] (o2) {\(p'\)};

			\diagram*{
				(i1) --[photon] (A) --[photon] (o1),
				(i2) --[fermion] (A) --[fermion] (o2),
			};
		\end{feynman}
	\end{tikzpicture}
	\caption{Diagram for the abstract 2-particle process considered}
        \label{fig:DiagramCFF}
\end{figure}

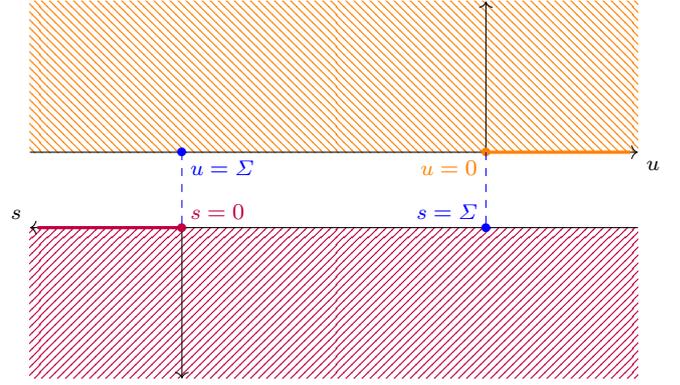
\begin{figure}[t]
	\centering
	\begin{tikzpicture}[scale=2]
		\fill[pattern=north west lines, pattern color=orange] (-1, 0) rectangle (3, 1);
		\fill[pattern=north east lines, pattern color=purple] (-1, -0.5) rectangle (3, -1.5);
		\draw[->] (-1, 0) -- (3, 0) node[anchor=north west] {$u$};
		\draw[->] (2, 0) -- (2, 1);
		\draw[->] (3, -0.5) -- (-1, -0.5)  node[anchor=south east] {$s$};
		\draw[->] (0, -0.5) -- (0, -1.5);
		\draw[orange, very thick] (2, 0) -- (2.95, 0);
		\node[orange] at (2,0) {$\bullet$};
		\node[orange, anchor=north east] at (2,0) {$u = 0$};
		\node[blue] at (0,0) {$\bullet$};
		\node[blue, anchor=north west] at (0,0) {$u = \Sigma$};
		\draw[purple, very thick] (-0.95, -0.5) -- (0, -0.5);
		\node[purple] at (0,-0.5) {$\bullet$};
		\node[purple, anchor=south west] at (0,-0.5) {$s = 0$};
		\node[blue] at (2,-0.5) {$\bullet$};
		\node[blue, anchor=south east] at (2,-0.5) {$s = \Sigma$};
		\draw[blue, dashed] (2, 0) -- (2, -0.5);
		\draw[blue, dashed] (0, 0) -- (0, -0.5);
	\end{tikzpicture}
	\caption{Representation of half-planes $\mathcal{P}_s^+$ and $\mathcal{P}_u^+$ aligned according to the crossing condition. Singularities for the $u$ and $s$ channel can appear only on the highlighted intervals.}
        \label{fig:Crossing}
\end{figure}

In order to derive an alternative expression to the DVCS dispersion relation at higer-pQCD order, we need to recall the analytic properties of the CFF, which are a special case of the  2-particle scattering amplitude (see fig. \ref{fig:DiagramCFF}).
These types of amplitudes are fully caracterised by the three Mandelstam variables $s$, $t$ and $u$ defined as :
\begin{align}
  \label{eq:Mandelstams}
  s & = (p +q )^2 = (p'+q')^2 \\
  \label{eq:Mandelstamt}
  t & = (p'-p)^2 = (q'-q)^2\\
  \label{eq:Mandelstamu}
  u & = (p'-q)^2 = (p-q')^2 
\end{align}
such that $s+t+u = p^2+q^2+q'^2+p'^2$.
In the following, we will assume working at fixed $t$ and fixed $q^2$ for a real outgoing photon, such that the amplitude can be fully described by either $s$ or $u$.
To distinguish between the two, we will write $\mathcal{F}_s(s)$ and $\mathcal{F}_u(u)$ respectively.
From that point, several postulates allow us to define an analytic continuation to the complex plane.
We briefly give them here but more details can be found in ref. \cite{Nussenzveig:1972tcd}.
\begin{enumerate}
\item Causality allows us to extend the physical amplitude to the upper-half complex plane $\mathcal{P}^+_s$, such that:
  \begin{equation}
    \label{eq:causality}
    \mathcal{F}_s(s) = \lim_{\epsilon \to 0} \mathcal{F}_s (s+i \epsilon).
  \end{equation}
\item The Schwartz reflection principle allows us to extend the upper plane analytic continuation to the lower one following:
  \begin{equation}
    \label{eq:SchwartzPrinciple}
    \mathcal{F}_s(s^*) = \mathcal{F}^*_s(s),
  \end{equation}
  and provided that $\mathcal{F}_s$ is real on at least a segment of the real axis.
  It immediately follows:
  \begin{align}
    \label{eq:RealPartCorollary}
    \lim_{\epsilon\to 0}\textrm{Re}\left(\mathcal{F}_s(s+i\epsilon)\right) & = \lim_{\epsilon\to 0}\textrm{Re}\left(\mathcal{F}_s(s-i\epsilon)\right), \\
    \label{eq:ImPartCorollary}
    \lim_{\epsilon\to 0}\textrm{Im}\left(\mathcal{F}_s(s+i\epsilon)\right) & = -\lim_{\epsilon\to 0}\textrm{Im}\left(\mathcal{F}_s(s-i\epsilon)\right),
  \end{align}
  and thus, $\mathcal{F}$ is discontinuous on the real axis if the imaginary part doesn't vanish.
\item The imaginary part of the amplitude can be computed from the optical theorem, and corresponds to the sum of all possible on-shell intermediate states.
  Single stable states are responsible for poles on the real axis, while multi-particle stable states trigger cuts.
  Importantly, if $s$, $t$ and $u$ are all space-like, then no on-shell intermediate state is allowed and the amplitude is real and continuous on the real axis. 
\end{enumerate}
Until now we have discussed the analytic continuation in the complex plane of $\mathcal{F}_s(s)$, but the same procedure could be applied to $\mathcal{F}_u(u)$.
The two variables are connected through:
\begin{align}
  \label{eq:suconnection}
  s+u = 2p^2 +q^2 -t = \Sigma
\end{align}
and therefore, $\mathcal{F}_s(s)$ and $\mathcal{F}_u(u)$ are expected to be related by the crossing symmetry and thus by analytical continuation.
However, in the case of CFFs, because one of the outgoing particles is massless (a real photon), to the best of our knowledge, there is no formal proof showing how to connect $\mathcal{F}_s(s)$ and $\mathcal{F}_u(u)$.
We therefore stick to the standard assumption stating that the analytic continuations of amplitudes $\mathcal{F}_s(s)$ and $\mathcal{F}_u(u)$ on their respective physical upper half-planes connect through a real interval between the two thresholds, as highlighted on fig. \ref{fig:Crossing}).

In the specific case of DVCS, we are able to easily identify such an interval.
Indeed, since $q^2$ is negative and much larger in absolute value than the hadron mass and $t$, $\Sigma$ is negative.
We can thus identify an interval between $(s=0,u=\Sigma)$ and $(s=\Sigma,u=0)$ where both $\mathcal{F}_s(s)$ and $\mathcal{F}_u(u)$ are real, allowing us to define the analytic continuation between the amplitudes.
We also highlight that in the case $\Sigma > 0$, the crossing structure is more complicated as the connection has to be done through the respective cuts of the amplitudes.
This situation is expected to arise in the case of Time-like Compton Scattering (TCS) where the hard scale is provided by a deeply timelike outgoing virtual photon.
We do not consider this case in the present analysis.

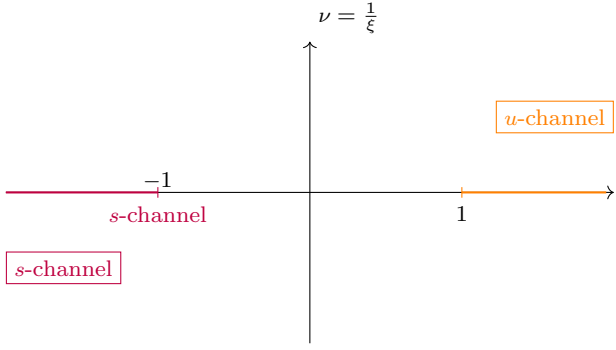
\begin{figure}[t]
	\centering
	\begin{tikzpicture}[scale=2]
		\draw (0, 1) node[anchor=south west] {$\nu = \frac{1}{\xi}$};
		\draw[->] (-2, 0) -- (2, 0);
		\draw[->] (0, -1) -- (0, 1);
		\draw[orange, thick] (1, 0) -- (1.95, 0);
		\draw[orange] (1, -1pt) -- (1, 1pt);
		\draw (1, -1pt) node[anchor=north]{$1$};
		\draw[purple, thick] (-1, 0) -- (-2, 0);
		\draw (-1, -1pt) node[anchor=south]{$-1$};
		\draw[purple] (-1, 1pt) -- (-1, -1pt);
		\draw[purple] (-1, -1pt) node[anchor=north]{$s$-channel};
		\node[draw, orange, anchor=east] at (2, 0.5) {$u$-channel};
		\node[draw, purple, anchor=west] at (-2, -0.5) {$s$-channel};
	\end{tikzpicture}
	\caption{Analytic structure of the amplitude in the $\nu$-plane}
        \label{fig:nuplane}
\end{figure}

The last step to characterise the properties of the CFF is to describe their singularity structure in the Bjorken limit\footnote{We neglect here complications coming from finite $t$ corrections \cite{Goldstein:2009ks}.}. We recall that in this limit, all masses, thresholds, and $|t|$ are very small compared to $Q^2 = -q^2$. Let us introduce the variable $\nu$ (differing by a factor 2 compare to the one of \cite{Diehl:2007jb})
\begin{equation}
  \label{eq:nuDef}
  \nu = \frac{s-u}{\Sigma} \overset{\mathrm{Bj}}{\approx} \frac{u-s}{Q^2} \overset{\mathrm{Bj}}{\approx} \frac{1}{\xi},
\end{equation}
such that for $\nu = 1$, $(s=\Sigma, u=0)$ and for $\nu=-1$, $(s=0,u=\Sigma)$. Consequently, the CFF $\mathcal{F}(\nu)$ is real and continuous for $\nu \in ]-1;1[$, and analytic for the entire complex plane but $\nu \in ]-\infty,-1] \cup [1,\infty[ = \mathcal{P}$. This structure is simple and illustrated on figure \ref{fig:nuplane}. The direct and crucial consequence is that for any $|\nu| < 1$, one can write the CFFs as:
\begin{align}
  \label{eq:CFFAnalytic}
  \mathcal{F}(\nu) = \sum_{j=0}^{\infty}f_j \nu^j 
\end{align}
or in terms of the variable $\xi$, for $|\xi|>1$:
\begin{align}
  \label{eq:CFFAnalyticXi}
  \mathcal{F}(\xi) = \sum_{j=0}^{\infty}f_j \frac{1}{\xi^j}. 
\end{align}
Note that because of the Schwartz principle, the $f_j$ are all real and uniquely define the analytic continuation in the complex plane. This generalises the proof provided of analyticity provided in ref. \cite{Teryaev:2005uj} which was based on a Taylor expansion of the LO DVCS kernel in the unphysical region $|\xi|>1$.

\subsection{All order dispersion relation with arbitrary subtraction}

The analyticity of the CFFs of eq. \eqref{eq:CFFAnalyticXi} is a major result, and our next goal is to connect the $f_j$ coefficients with the associated GPDs. To derive these relations, we first go back to eq. \eqref{eq:CFFquarkDef}, where the factorisation theorem is applied to write the CFF as a convolution of coefficient function $T$ and a GPD $H$. The $\xi$ dependance in these formula can be simplified by using the Double Distribution representation of GPDs. Thus introducing eq. \eqref{eq:DDF} into \eqref{eq:CFFquarkDef}, one obtains:
\begin{align}
  \label{eq:CFFintermsofDD}
  \mathcal{H}^q = \frac{1}{\xi}\int_{\Omega}T^q\left(\alpha+\frac{\beta}{\xi}\right) F^q(\beta, \alpha) \,\ud\beta\ud\alpha + h_0,
\end{align}
where we define:
\begin{align}
  \label{eq:subtractionDD}
  h_0 = \int_{-1}^1T^q(\omega)D^q(\omega)\,\ud\omega.
\end{align}

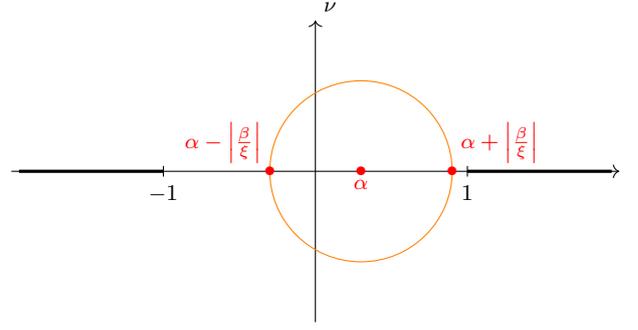
\begin{figure}[t]
  \centering
  \begin{tikzpicture}[scale=2]
    \draw[->] (-2, 0) -- (2, 0);
    \draw[->] (0, -1) -- (0, 1) node[anchor=south west] {$\nu$};
    \draw (1, 1pt) -- (1, -1pt) node[anchor=north] {$1$};
    \draw[very thick] (1, 0) -- (1.95, 0);
    \draw (-1, 1pt) -- (-1, -1pt) node[anchor=north] {$-1$};
    \draw[very thick] (-1, 0) -- (-1.95, 0);
    \draw[red] (0.3, 0) node {$\bullet$} node[anchor=north] {$\alpha$};
    \draw[orange] (0.3, 0) circle (0.6);
    \draw[red] (0.9, 0) node {$\bullet$} node[anchor=south west] {$\alpha+\left|\frac{\beta}{\xi}\right|$};
    \draw[red] (-0.3, 0) node {$\bullet$} node[anchor=south east] {$\alpha-\left|\frac{\beta}{\xi}\right|$};
  \end{tikzpicture}
  \caption{The range of the argument inside function $T$}
  \label{fig:AnalyticityofT}
\end{figure}

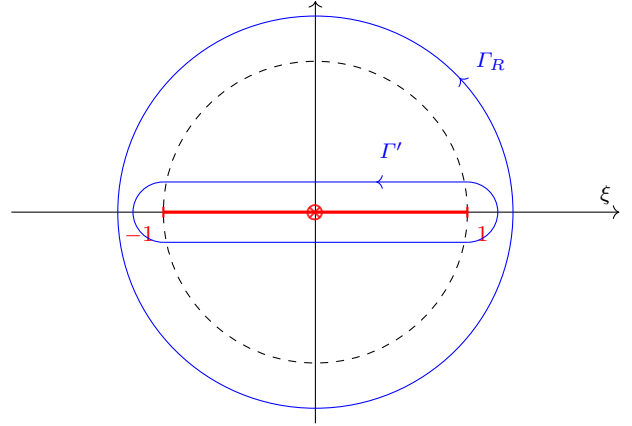
\begin{figure}[t]
  \centering
  \begin{tikzpicture}[scale=2]
    \draw[->] (-2, 0) -- (2, 0) node[anchor=south east] {$\xi$};
    \draw[->] (0, -1.4) -- (0, 1.4);
    \draw[dashed] (0, 0) circle (1);
    \draw[red] (0, 0) node{$\pmb{\otimes}$};
    \draw[red, thick] (1, 1pt) -- (1, -1pt) node[anchor=north west] {$1$};
    \draw[red, very thick] (-1, 0) -- (1, 0);
    \draw[red, thick] (-1, 1pt) -- (-1, -1pt) node[anchor=north east] {$-1$};
    \draw[blue,
    decoration={markings, mark=at position 0.12 with {\arrow{>}}},
    postaction={decorate}
    ] (0, 0) circle (1.3);
    \draw[blue] (-1, -0.2) arc (-90:-270:0.2);
    \draw[blue,
    decoration={markings, mark=at position 0.3 with {\arrow{>}}},
    postaction={decorate}
    ] (1, 0.2) -- (-1, 0.2);
    \draw[blue] (-1, -0.2) -- (1, -0.2);
    \draw[blue] (1, -0.2) arc (-90:90:0.2);
    \draw[blue] (1, 1) node[anchor=west] {$\Gamma_R$};
    \draw[blue] (0.5, 0.3) node[anchor=south] {$\Gamma'$};
  \end{tikzpicture}
  \caption{The contours used in the proof of dispersion relations. In red, the singularities and branch cuts.}
  \label{fig:Contours}
\end{figure}

\noindent Note that we have assumed that the Radon transform can be permutted with the convolution over $x$. From eq. \eqref{eq:CFFintermsofDD}, one sees that the analytic properties of $\mathcal{H}$ are a direct consequence of the analytic properties of $T$. And in fact these properties are the same, since both $\mathcal{H}$ and $T$ describe the scattering of two particles. $T(z)$ is analytic for $|z|< 1$ and in our case for $|\xi| > 1$:
\begin{align}
  \label{eq:OrderingTargument}
  -1 < \alpha - \left|\frac{\beta}{\xi}\right| \le \alpha +\frac{\beta}{\xi}\le \alpha + \left|\frac{\beta}{\xi}\right| < 1 ,
\end{align}
as the support of the DD is limited to $\{(\alpha,\beta)| |\alpha|+ |\beta|\le 1\}$. Consequently, for every value of $\alpha$ in the DD support, $T(\alpha + \beta/\xi)$ is analytic in the unphysical region (see figure \ref{fig:AnalyticityofT}), and thus Taylor expanded around $\alpha$ into:
\begin{align}
  \label{eq:TTaylorExpand}
  T\left(\alpha+\frac{\beta}{\xi}\right) = \sum_{n=0}^\infty \frac{1}{n!} \frac{\partial^n T}{(\partial \alpha)^n}(\alpha) \left(\frac{\beta}{\xi}\right)^n, \quad \textrm{for } |\xi| >1.
\end{align}
Injecting eq. \eqref{eq:TTaylorExpand} into \eqref{eq:CFFintermsofDD}, we can compute the coefficients introduced in eq. \eqref{eq:CFFAnalyticXi} in terms of moments of the DD $f$ and derivatives of the coefficient function $T$:

\begin{align}
  \label{eq:Hexpansion}
  \mathcal{H}(\xi) & = \sum_{j=0}^\infty h_j \frac{1}{\xi^j} \quad \textrm{for } |\xi| >1,\\
  \label{eq:Defh0}
  h_0 & = \int_{-1}^1  T^q(\omega) D^q(\omega)\,\ud\omega \\
  \label{eq:Defhj}
  h_{j+1} & =  \frac{1}{j!}\int_{-1}^1\textrm{d}\alpha \frac{\partial^j T}{(\partial \alpha)^j}(\alpha) \int_{-1+|\alpha|}^{1-|\alpha|}\beta^j F^q(\beta,\alpha)\,\ud \beta
\end{align}
\noindent This completes our first goal, but these results are restricted to the unphysical region. One needs to use the dispersion relation to derive relations between these coefficients in the physical region.
To do this, we define the following $\mathcal{I}_n$ integrals in the complex plane:
\begin{align}
  \label{eq:CircleIntegral}
  \mathcal{I}_n(\xi)  & = \oint_{\Gamma_R} \frac{\mathcal{H}(\xi')}{\xi'-\xi}\left(\frac{\xi'}{\xi}\right)^n\,\ud\xi' \nonumber \\
                      & = \sum_j h_j \oint \frac{\xi'^{-j}}{\xi'-\xi}\left(\frac{\xi'}{\xi}\right)^n\,\ud\xi',
\end{align}%
\noindent where the contour $\Gamma_R$ is illustrated on fig. \ref{fig:Contours} and chosen such that the CFF is analytic all along. Note that in eq. \eqref{eq:CircleIntegral}, $\xi'$ is in the unphysical region, allowing us to expand the CFF $\mathcal{H}$, but $\xi$ can be safely choosen in the physical region. Indeed, taking $\xi \in [-1,1]$, one gets:
\begin{align}
  \label{eq:ResidueTheorem}
  \oint_{\Gamma_R} \frac{\xi'^{-k}}{\xi'-\xi} \,\ud\xi' =\begin{cases}
    0 & \textrm{if}\ k > 0 \\
    2i\pi\xi^{-k} & \textrm{if}\ k \le 0
    \end{cases} ,
\end{align}
as in the case $k > 0$, the contributions of the two poles exactly compensate each other. We deduce:
\begin{align}
  \label{eq:InResult}
   \mathcal{I}_n(\xi)  & = 2\pi i \sum_{j=0}^{n} h_j \frac{1}{\xi^j},
\end{align}
showing that the $\mathcal{I}_n$ truncate eq. \eqref{eq:Hexpansion} to order $n$.

The next step is to connect $\mathcal{I}_n$ with the value of the CFF within the physical region.
To do so, we can deform the contour integration from $\Gamma_R$ to $\Gamma'$ such that:

\vspace{2cm}
\begin{strip}
  \rule[-1ex]{\columnwidth}{1pt}\rule[-1ex]{1pt}{1.5ex}
  \begin{align}
    \label{eq:INPhysical}
    \mathcal{I}_n(\xi) & = \int_{-1}^{1}\frac{\mathcal{H}(\xi'-i\varepsilon)}{\xi'-\xi-i\varepsilon}\left(\frac{\xi'}{\xi}\right)^n\,\ud\xi' - \int_{-1}^{1}\frac{\mathcal{H}(\xi'+i\varepsilon)}{\xi'-\xi+i\varepsilon}\left(\frac{\xi'}{\xi}\right)^n\,\ud\xi'  \nonumber \\
                       &= \fint_{-1}^{1}\frac{\mathcal{H}(\xi'-i\varepsilon)}{\xi'-\xi}\left(\frac{\xi'}{\xi}\right)^n\,\ud\xi' + i\pi\mathcal{H}(\xi-i\epsilon) - \fint_{-1}^{1}\frac{\mathcal{H}(\xi'+i\varepsilon)}{\xi'-\xi}\left(\frac{\xi'}{\xi}\right)^n\,\ud\xi'+i\pi\mathcal{H}(\xi+i\epsilon)
  \end{align}
   \hfill\rule[1ex]{1pt}{1.5ex}\rule[2.3ex]{\columnwidth}{1pt}
\end{strip}%
\noindent where we used the Sokhotski–Plemelj formula. The Schwartz reflexion principle in eq. \eqref{eq:SchwartzPrinciple} allows us to rewrite the formula in terms real and imaginary parts of the CFF such that:
\begin{align}
  \label{eq:INPhysical2}
  \mathcal{I}_n(\xi) = 2i\pi \Re \mathcal{H}(\xi) + 2i \fint_{-1}^{1}\frac{\Im\mathcal{H}(\xi')}{\xi'-\xi}\left(\frac{\xi'}{\xi}\right)^n\,\ud\xi'
\end{align}
after safely taking the limit $\epsilon \to 0^+$. An additional subtlety comes from the fact that $\Im[\mathcal{H}]$ is defined as the limit coming from the upper half-plane (see eq. \eqref{eq:causality}) using Mandelstam variables, and thus $\nu$. Since $\xi = 1/\nu$, we have:
\begin{align}
  \label{eq:XilowerPlane}
  \Im \mathcal{H} = \lim_{\epsilon\to 0^+} \mathcal{H}(\nu+i\epsilon) = \lim_{\epsilon\to 0^+}\mathcal{H}(\xi - i\epsilon),
\end{align}
triggering a plus sign in front of integral.

One should realise that the result in eq. \eqref{eq:INPhysical2} is conditioned to the behaviour of $\mathcal{H}$ within the integration contour, and in particular for $\xi \to 0$.
    Indeed the contour deformation of eq. \eqref{eq:INPhysical} can be performed only if the integrand remain integrable on the contour, especially for $\xi \to 0$ (or equivalently $\nu \to \infty$).
    We expect the CFF to present a Regge behaviour in $\xi^{-\alpha}$ for $\xi \to 0$.
Then expressions \eqref{eq:INPhysical} and  \eqref{eq:INPhysical2} for  $\mathcal{I}_n$ are only valid for $n>\alpha -1$.

Combining eqs. \eqref{eq:InResult} and \eqref{eq:INPhysical2} we can deduce the general expression for the $n$-times subtracted dispersion relation at any order of perturbation theory:
\begin{align}
  \label{eq:DispersionRelation}
   \Re \mathcal{H}(\xi) +\frac{1}{\pi} \fint_{-1}^{1}\frac{\Im\mathcal{H}(\xi')}{\xi'-\xi}\left(\frac{\xi'}{\xi}\right)^n\,\ud\xi' = \sum_{j=0}^{n} h_j \frac{1}{\xi^j},
\end{align}
with $h_j$ being given in eqs. \eqref{eq:Defh0} and \eqref{eq:Defhj}. It is easy to verify that $h_0$ is in fact the subtraction constant $S^q$ we introduced in eq. \eqref{eq:SubtractionLO}.

\subsection{New expression for the subtraction constant and consistency with previous results}

Eq. \eqref{eq:DispersionRelation} is the key result of this section allowing us to connect the real and imaginary parts of the CFF to the D-term and the Double Distribution. Yet, it can be simplified using the symmetries of DVCS and GPDs. Importantly, the CFFs are $\xi$ even and thus $\mathcal{H}(\xi) = \mathcal{H}(-\xi)$. When combining this parity argument with the Schwarz reflection principle $\mathcal{H}(\xi^*) = \mathcal{H}^*(\xi)$ we obtain the following constraint:
\begin{align}
  \label{eq:SymmetryCFF}
  \mathcal{H}(-\xi^*) = \mathcal{H}^*(\xi).
\end{align}
From this we deduce that the real part of the CFF must be even in $\xi$, while the imaginary one must be odd along the real axis, \emph{i.e.} as a function of the real part of $\xi$. This restricts our expansion in eq. \eqref{eq:Hexpansion} to
\begin{align}
    \label{eq:HexpansionSymmetry}
  \mathcal{H}(\xi) & = \sum_{j \textrm{ even}}^\infty h_j \frac{1}{\xi^j}.
\end{align}
As a consequence, restricting ourselves to $\xi \in ]0,1[$, the dispersion relation \eqref{eq:DispersionRelation} can be simplified into:
\begin{align}
    \label{eq:SymDispersionRelation}
&\sum_{j\text{ even}}^nh_j\xi^{-j} =  \Re \mathcal{H}(\xi) \nonumber \\
&-\frac{1}{\pi}\fint_{0}^1\Im \mathcal{H}(x)\left(\frac{x}{\xi}\right)^n\bigg[\frac{1}{\xi-x} - (-1)^n \frac{1}{\xi+x}\bigg]\,\ud x \\
&= \Re \mathcal{H}(\xi) - \frac{2}{\pi}\fint_{0}^1\left(\frac{x}{\xi}\right)^{n}\frac{x \Im \mathcal{H}(x)}{(\xi-x)(\xi+x)}\,\ud x
\end{align}
where $k = 2 [\frac{n}{2}]$ is the largest even number inferior or equal to $n$. $n = 0$ and $n = 1$ are therefore equal, and exactly identical to the usual formula in eq. \eqref{eq:SubtractionLO}. For phenomenological CFFs, for which $x \Im \mathcal{H}(x)$ is integrable, this expression converges. Note that the terminology can be misleading. $\mathcal{S}$ is usually called the subtraction constant, while we show here that it is actually extracted from an unsubtracted dispersion relation.

On top of eq. \eqref{eq:subtractionDD} for the quarks contribution, the above discussion can be generalised to the gluons contribution with:
\begin{equation}
  \label{eq:subtractionDDg}
  \mathcal{S}^g = \int_{-1}^1 T^{g}(\omega) D^g(\omega) \textrm{d}\omega .
\end{equation}
Because of the Schwartz principle, $h_0$ (and in fact all the $h_j$) is real, which means that only the real part of $T^{q,g}$ contribute to eq. \eqref{eq:subtractionDD} and \eqref{eq:subtractionDDg}.
  We are thus left with the following results:
  \begin{equation}
    \label{eq:SubtractionDFinal}
    \mathcal{S}^a = \int_{-1}^1 \Re T^{a}(\omega) D^a(\omega) \textrm{d}\omega .
  \end{equation}

Eq. \eqref{eq:SubtractionDFinal} can be recovered from the results presented in ref. \cite{Diehl:2007jb} and recalled in \eqref{eq:DiehlDR}. There, the subtraction constant is expressed as a double convolution involving the $D$-term and the imaginary part of the perturbative kernel. Reshuffling the expression using the odd-parity of the $D$-term:
  \begin{align}
    \label{eq:DiehlIvanovSubtraction}
    \mathcal{S}^q = &\frac{2}{\pi} \int_1^\infty \textrm{d}\omega \Im T^q(\omega) \int_{-1}^1 \textrm{d} \alpha \frac{D^q(\alpha)}{\omega-\alpha} \nonumber \\
    = & \int_{-1}^1 \textrm{d} \alpha D^q(\alpha) \frac{1}{\pi} \int_1^\infty \textrm{d}\omega \Im T^q(\omega)  \left[\frac{1}{\omega-\alpha} - \frac{1}{\omega + \alpha} \right]        
  \end{align}
  and injecting now the dispersion relation of the hard scattering kernel obtained in \cite{Diehl:2007jb} :
  \begin{equation}
    \label{eq:DiehlIvanovTDR}
    \Re T^q(\nu) = \frac{1}{\pi} \int_1^\infty \textrm{d}\omega \Im T^q(\omega)\left[\frac{1}{\omega-\nu}-\frac{1}{\omega+\nu} \right],
  \end{equation}
  we recover our eq. \eqref{eq:SubtractionDFinal}.

\subsection{Higher subtractions and connection with the DDs}

An unexpected result of this new derivation of the higher-order connection between the subtraction constant and the $D$-term, is that the dispersion relation formalism allows us to connect the Mellin moments of the Double Distribution with higher-order subtraction constants.
Indeed, going back to eq. \eqref{eq:Hexpansion}-\eqref{eq:Defhj}, one realises that we have only exploited the connection provided by eq. \eqref{eq:Defh0}.
It is possible to isolate the $h_j$ for $j\neq 0$ by subtracting two consecutive terms in eq. \eqref{eq:SymDispersionRelation} (we recall that $\xi \in ]0,1[$):
\begin{align}
  \label{eq:HigherDR}
  h_n \xi^{-n} = & \frac{1}{\pi}\fint_{0}^1\frac{\Im \mathcal{H}(\xi')}{\xi'-\xi}\left(\frac{\xi'}{\xi}\right)^{n-1}\left(\frac{\xi'}{\xi}-1\right)\,\ud\xi' \nonumber \\
  &  +  \frac{(-1)^n}{\pi}\int_{0}^1\frac{\Im \mathcal{H}(\xi')}{\xi'+\xi}\left(\frac{\xi'}{\xi}\right)^{n-1}\left(\frac{\xi'}{\xi}+1\right)\,\ud\xi'.
\end{align}
After simplification, this can be simply written for $j \geq 1$ as:
\begin{align}
  \label{eq:HigherDR}
  h_{2j} = & \frac{2}{\pi}\int_{0}^1\Im \mathcal{H}(\xi')\left(\xi'\right)^{2j-1}\,\ud\xi'.
\end{align}
Reinjecting eq. \eqref{eq:Defhj}, we get for $\ell$ odd:
\begin{align}
  \label{eq:FinalHigherOrder}
  & \frac{2}{\pi}\int_{0}^1\Im \mathcal{H}(\xi')\left(\xi'\right)^{\ell}\,\ud\xi' \nonumber \\
= & \frac{1}{\ell!}\int_{-1}^1\textrm{d}\alpha \frac{\partial^{\ell} T}{(\partial \alpha)^{\ell}}(\alpha) \int_{-1+|\alpha|}^{1-|\alpha|}\textrm{d}\beta \beta^{\ell} f^q(\beta,\alpha)
\end{align}
Rewording this equation, we note that the $\ell^{\textrm{th}}$ Mellin moment of the imaginary part of the CFF is connected with the $\beta$-moment of the Double Distribution, convoluted then with the derivative of the hard scaterring kernel. This equation highlights well the deconvolution problem of GPDs and DDs from DVCS data \cite{Bertone:2021yyz}.
Indeed, while two indices are required to independently deconvolute the $(\alpha,\beta)$ dependence of the DDs, only a single index, $\ell$, appears here.
The impact of such relation on the caracterisation of shadow GPDs is left for future work.

Finally, let us mention that the fact of implementing dispersion relation at the level of the CFF directly may have a significant impact already at the level of CFF extraction (see figure 2 of \cite{Cuic:2020iwt} for an illustration of the impact of the one-subtracted dispersion relation for DVCS). Assessing the impact of higher-subtracted dispersion relations on CFF extraction is also left for future studies.


%% file: section4.tex
If we limit ourselves to the lowest subtraction, which only involves the $D$-term, we have demonstrated that the CFFs give us in principle access to the subtraction constant:
\begin{equation}
    \mathcal{S}^a(t, Q^2) = \int_{-1}^1 \ud \omega\,\Re T^a\left(\omega, \frac{Q^2}{\mu^2}, \alpha_s\right) D^a(\omega, t, \mu^2)\,. 
\end{equation}
We will elaborate on the challenges related to the characterization of $S^a$ from experimental data in the next section. For now, we are interested in the fact that the GFF $C_a(t, \mu^2)$, which is related to the pressure distribution in the proton, writes as another integral of the $D$-term, eq. \eqref{eq:eihwjcnalkxm} which we recall with its full variable dependence:
\begin{equation}
    C_a(t, \mu^2) = \frac{1}{4}\int_{-1}^1 \ud \alpha\,\alpha^{1-p_a} D^a(\alpha, t, \mu^2)\,.
\end{equation}
An obvious question is whether the knowledge of the subtraction constant $S^a(t, Q^2)$ allows an unambiguous extraction of the GFF $C_a(t, \mu^2)$. This question is very similar to that known as the deconvolution problem \cite{Bertone:2021yyz}, which aims at determining whether the measurement of CFFs -- \textit{i.e.} the convolution of the perturbative coefficient function to the full GPD -- allows the unambiguous reconstruction of the GPD. In fact, the problem at hand in this paper is exactly the restriction of the general deconvolution problem to the $D$-term. 

As we have already hinted at, the root of the deconvolution problem is that DVCS experimental data offers one less kinematic variable compared to the parton distributions we want to extract. CFFs are functions of $(\xi, t, Q^2)$, whereas GPDs are functions of $(x, \xi, t, \mu^2)$; the subtraction constant is a function of $(t, Q^2)$ whereas the $D$-term of $(\alpha, t, \mu^2)$. One might argue that the GFF $C_a$ which we are fundamentally interested in, is just a function of $(t, \mu^2)$ -- so a similar kinematic dependence as the subtraction constant $S^a(t, Q^2)$. However, one cannot write a straightforward relation between the subtraction constant $S^a(t, Q^2)$ and the GFF $C_a(t, \mu^2)$ which does not involve in practice the extraction of the $D$-term $D^a(\alpha, t, \mu^2)$. 

There is however a theoretical solution to the missing variable problem. At a given order in perturbation theory, the scale dependence of $D^a(\alpha, t, \mu^2)$ is given by renormalization group equations, removing in principle one degree of freedom. In practice, evolution equations entangle the $(\alpha, \mu^2)$ dependence of the $D$-term (and the $(x, \xi, \mu^2)$ dependence of GPDs). However, although this solves the issue on paper, in practice, effect of QCD evolution are rather weak on the range of $Q^2$ accessible to exclusive processes. Ref. \cite{Bertone:2021yyz} offered an explicit construction of very different GPDs (with vanishing $D$-terms) such that their CFFs would be indiscernable to experimental data. These shadow GPDs represent an illustration of particularly badly constrained objects to DVCS in any foreseeable data. Solutions to this issue involve, on the one hand, the introduction of more theoretical constraints to reduce the functional space accessible to GPDs \cite{Dutrieux:2021wll}, and on the other hand an ambitious program of global fits on a variety of exclusive processes. In particular, processes which do not show an exacerbated sensitivity to a pole $x = \xi$ as DVCS, TCS or DVMP, but rather to a pole where $x$ and $\xi$ are entangled to an external kinematic variable are very desirable, like DDVCS \cite{Guidal:2002kt,Belitsky:2003fj,Deja:2023ahc} or two-to-three exclusive processes \cite{Duplancic:2018bum,Pedrak:2020mfm,Grocholski:2021man,Grocholski:2022rqj,Qiu:2022bpq,Qiu:2022pla,Duplancic:2023kwe,Qiu:2023mrm}. 

In a similar fashion to the study of the deconvolution problem led in Ref. \cite{Bertone:2021yyz}, there exist shadow $D$-terms, which bring barely any contribution to the subtraction constant over current ranges in $Q^2$ and are therefore extremely hard to discern in the data. Ref. \cite{Dutrieux:2021nlz} gave a hint at such shadow $D$-term when it highlighted a tremendous increase of uncertainty as soon as the parametrization of the $D$-term was made slightly more flexible. In practice, it is common to parametrize the $D$-term through an expansion in Gegenbauer moments due to their friendly LO evolution properties:
\begin{align}
    D^q(\alpha, t, \mu^2) &= (1-\alpha^2) \sum_{\textrm{odd}\ n} d_n^q(t, \mu^2) C^{(3/2)}_n(\alpha)\,,\\
    D^g(\alpha, t, \mu^2) &= \frac{3}{2}(1-\alpha^2)^2 \sum_{\textrm{odd}\ n} d_n^g(t, \mu^2) C^{(5/2)}_{n-1}(\alpha)\,.
    \label{eq:Gegen}
\end{align}
As Gegenbauer polynomials form a complete orthogonal family, this representation is fairly general -- but comes with the drawback that fixed-order truncations are usually oscillating functions. We refer to Ref. \cite{Dutrieux:2021nlz} for an account of the LO scale dependence of $d_n^a(t, \mu^2)$. In this representation, the LO subtraction constant reads as:
\begin{equation}
\mathcal{S}(t, Q^2) \eqLO 4\sum_q e_q^2 \sum_{\textrm{odd}\ n} d_n^q(t, \mu^2)\,, \label{eq:LOsubGegen}
\end{equation}
where we will use conventionally in the following $\mu^2 \equiv Q^2$. On the other hand, the GFF $C_a(t, \mu^2)$ reads:
\begin{equation}
    C_a(t, \mu^2) = \frac{1}{5}\,d^a_1(t, \mu^2)\,.
\end{equation}
The problem of relating the subtraction constant to the pressure inside the proton turns into the question of extracting $d_1^a$ from the sum of all $d_n^a$ at LO (and more complicated infinite linear combinations of the $d_n^a$ at higher order). A simple solution to the ill-definedness of this extraction is to assume that only a finite number of coefficients $d_n^a$ actually contribute to the subtraction constant. In fact, the study of \cite{Burkert:2018bqq} used only $n = 1$, and evaluated the systematic uncertainty caused by such a rigid modelling of the $D$-term by inputs from a quark soliton model. In \cite{Dutrieux:2021nlz}, effects of a truncation at $n= 1$ and $n=3$ were compared. It was observed that the -- already large -- uncertainty on $d_1$ inflated by a factor 20 when $d_3$ was allowed to be non-zero. In fact, the reason is fairly simple to understand. Since evolution effects are relatively small on the narrow range in $Q^2$ available to the current precise DVCS data, $d_1(t, \mu^2)$ and $d_3(t, \mu^2)$ do not exhibit a significantly different behavior in $\mu^2$. Therefore, parasitic contributions such that
\begin{equation}
    d_1^q(t, \mu^2) \approx -d_3^q(t, \mu^2)
\end{equation}
amount to almost no contribution to the LO subtraction constant of eq. \eqref{eq:LOsubGegen}, and are virtually unconstrained. An object which brings exactly no contribution to the subtraction constant at a given scale $\mu_0^2$ will be called a \textit{shadow} $D$-term, and we have already highlighted a very simple example at LO:
\begin{equation}
    d_1^q(\mu_0^2) = \lambda\,; \ d_3^q(\mu_0^2) = -\lambda\,,\label{eq:fevhbksnw}
\end{equation}
or equivalently
\begin{equation}
    D_{S,LO}^q(\alpha, \mu_0^2) = \lambda(1-\alpha^2)[C_1^{(3/2)}(\alpha) - C_3^{(3/2)}(\alpha)]\,.
\end{equation}
The space of shadow $D$-terms at a fixed scale is a vector space (it is the kernel, or null-space of the integral transform and there exist shadow $D$-terms of arbitrary size). Under evolution to another scale $\mu^2 \neq \mu_0^2$, the contribution of a shadow $D$-term to the subtraction constant becomes non-zero. Indeed, \eqref{eq:fevhbksnw} can only be true at one scale, since the $\mu^2$ dependence of both sides of the equation are ruled by different anomalous dimensions. Therefore, the range in scales on which DVCS is measured precisely directly constrains the maximal size of shadow $D$-terms, and the uncertainty of the deconvolution procedure.

To give a simple approximate example, if there were no mixing between quarks and gluons, the evolution of $d_n^q$ would be entirely dictated by the anomalous dimension $\gamma_n$ following
\begin{align}
&d_n^q(\mu^2) = \Gamma_n^{qq}(\mu^2, \mu_0^2) d_n^q(\mu_0^2)\,,\nonumber \\ 
&\hspace{20pt}\textrm{ where }\Gamma_n^{qq}(\mu^2, \mu_0^2) = \left(\frac{\alpha_s(\mu^2)}{\alpha_s(\mu_0^2)}\right)^{2\gamma_n / \beta_0}\,,
\end{align}
where $\beta_0$ is the first coefficient in the $\beta$ function of $\alpha_s$. Using $\gamma_1 = 16/9$, $\gamma_3 = 157/45$ and $\beta_0 = 11 - 2 n_f / 3 = 9$, we find that the contribution to the subtraction constant of the simple shadow $D$-term of eq. \eqref{eq:fevhbksnw} is:
\begin{align}
\mathcal{S}_S^q(Q^2) &= \frac{8}{3}(\Gamma_1^{qq}(Q^2, \mu_0^2) d_1^q(\mu_0^2) + \Gamma_3^{qq}(Q^2, \mu_0^2) d_3^q(\mu_0^2)) \,,\\
&\approx \frac{8}{3} \lambda \bigg[\left(\frac{\alpha_s(Q^2)}{\alpha_s(\mu_0^2)}\right)^{0.395}-\left(\frac{\alpha_s(Q^2)}{\alpha_s(\mu_0^2)}\right)^{0.775}\bigg]\,.
\end{align}
This gives of course 0 if $Q^2 = \mu_0^2$ by definition of the shadow $D$-term. Linearizing the previous relation yields this approximate contribution of the shadow $D$-term to the subtraction constant: 
\begin{align}
\mathcal{S}_S^q(Q^2) \approx  \lambda  \bigg[1-\frac{\alpha_s(Q^2)}{\alpha_s(\mu_0^2)}\bigg]\,.
\end{align}
If the experimental uncertainty of the subtraction constant is characterized by a quantity $\Delta S$, and the measurement have been performed on a range in scales of $[Q^2_{min}, Q^2_{max}]$, our approximate approach tells that the shadow $D$-term of eq. \eqref{eq:fevhbksnw} will be typically bring a dispersion:
\begin{align}
\sigma_{S,d1q} &\approx \sigma_{S,d3q} \nonumber \\&\approx \frac{3}{8} \frac{\Delta S}{\Gamma_1^{qq}(Q^2_{max}, Q^2_{min}) - \Gamma_3^{qq}(Q^2_{max}, Q^2_{min})}  \label{eq:backenvest_ex}\\
&\approx \frac{\Delta S}{\displaystyle \left(1-\frac{\alpha_s(Q_{max}^2)}{\alpha_s(Q^2_{min})}\right)}\,. \label{eq:backenvest}
\end{align}
The approximate form of the last line should only be used for scales close to the charm mass, as it is derived with the anomalous dimensions of $n_f = 3$. Eq. \eqref{eq:backenvest_ex} is general on the other hand, provided the true evolution operator is used. Of course, there exist many more shadow $D$-terms if the parametrization in terms of $d_n$ is made more flexible than solely $d_1^q$ and $d_3^q$, in particular if explicit gluons contributions are included. The simple eq. \eqref{eq:backenvest} represents a typical estimate of the uncertainty of the deconvolution procedure within the parametric space which we have chosen. Despite the simplifying assumptions that we have made, this result captures the essence of the propagation of uncertainty: first a dependence on the experimental uncertainty of the data through $\Delta S$, and then a characterization of how different the evolution of the different parameters is with  respect to the scale. 

Formally, since the fit of the $d_n^a$ coefficients is linear, it is straightforward to write the exact solution of the fit. In particular, the covariance matrix of $d_1^q$ and $d_3^q$ that we are interested in for this simple example writes:
\begin{equation}
\begin{pmatrix}
\sigma_{d1q}^2 & \textrm{cov}[d_1^q, d_3^q] \\
\textrm{cov}[d_1^q, d_3^q] & \sigma_{d3q}^2
\end{pmatrix} = (C^T \Omega^{-1} C)^{-1}\,, \label{eq:exact}
\end{equation}
where $\Omega$ is the covariance matrix of the fitted dataset and $C$ is the so-called design matrix which contains the values of the fitted functions at the fitted kinematics, here:
\begin{equation}
C = \frac{8}{3}\begin{pmatrix}
\Gamma_1^{qq}(Q_1^2, \mu_0^2) & \Gamma_3^{qq}(Q_1^2, \mu_0^2) \\
\Gamma_1^{qq}(Q_2^2, \mu_0^2) & \Gamma_3^{qq}(Q_2^2, \mu_0^2) \\
\vdots
\end{pmatrix}\,. 
\end{equation}
One can draw a parallel between this exact general formula and the approximate shadow $D$-term uncertainty that we have derived in eq. 
\eqref{eq:backenvest}. If there were only two measurements in $Q^2$, one at $Q^2_{min}$ and one at $Q^2_{max}$, we chose to evaluate $\mu_0^2 = Q^2_{min}$, and the experimental dataset was uncorrelated with standard deviation $\Delta S$, we would find:
\begin{equation}
\sigma_{d1q} = \frac{3}{8}\Delta S \frac{\sqrt{1 + [\Gamma_3^{qq}(Q^2_{max}, Q^2_{min})]^2}}{\Gamma_1^{qq}(Q^2_{max}, Q^2_{min}) - \Gamma_3^{qq}(Q^2_{max}, Q^2_{min})} \label{eq:acds1}
\end{equation}
\begin{equation}
\sigma_{d3q} = \frac{3}{8}\Delta S \frac{\sqrt{1 + [\Gamma_1^{qq}(Q^2_{max}, Q^2_{min})]^2}}{\Gamma_1^{qq}(Q^2_{max}, Q^2_{min}) - \Gamma_3^{qq}(Q^2_{max}, Q^2_{min})} \label{eq:acds2}
\end{equation}
Using that $\Gamma^{qq}(Q^2_{max}, Q^2_{min}) < 1$, we find a very similar estimate to the one we have derived using the notion of shadow $D$-term, without needing the concept at all. Shadow $D$-terms are after all merely an attempt at simplifying, or making more intuitive, the analysis of the inverse linear problem by identifying obvious directions that are dominant in the uncertainty propagation. It is truly useful when it comes to making broad predictions based on general characteristics of the data without going through the process of generating pseudo-data and fitting them. We show such exercise to broadly evaluate the plausible impact of the EIC in the last section of this paper, where we will construct NLO shadow $D$-terms and treat the evolution equations properly.

Let us note in passing that, with the same assumptions that were used to derive eqs. \eqref{eq:acds1}-\eqref{eq:acds2}, we find:
\begin{equation}
\textrm{corr}[d_1^q, d_3^q] = \frac{-1-\Gamma_1^{qq} \Gamma_3^{qq}}{\sqrt{(1+[\Gamma_1^{qq}]^2)(1+[\Gamma_3^{qq}]^2)}}\,,\label{eq:correst}
\end{equation}
where we have omitted the argument $(Q^2_{max}, Q^2_{min})$. If evolution is very weak, $\Gamma_1^{qq} \approx \Gamma_3^{qq} \approx 1$, which gives as expected $\textrm{corr}[d_1^q, d_3^q] \approx -1$, with $\sigma_{d1q} \approx \sigma_{d3q} \approx +\infty$.


%% file: section5.tex
We now conduct a re-analysis of the LO extraction of the GFF $C_a(t)$ of Ref. \cite{Dutrieux:2021nlz} using the NLO DVCS coefficient function, and putting to use the understanding of the deconvolution uncertainty stemming from shadow $D$-terms. A neural network analysis of the global DVCS dataset was conducted in 2019 \cite{Moutarde:2019tqa}, leveraging 30 observables over 2500 kinematic configurations acquired during 17 years of measurements. The real and imaginary parts of the four leading-twist CFFs were modelled independently. The result of the fit was 100 sets of CFFs which represent a sample of the functional distribution of the CFFs. 

The computation of the subtraction constant from eq. \eqref{eq:SubtractionLO} requires the evaluation of the imaginary part of the CFF on the full range of $\xi \in ]0,1[$. Since the skewness $\xi$ is related to the plus component of the four-momentum transfer $\Delta$, it is bound kinematically by the value of $t$ according to:
\begin{equation}
|\xi| \leq \frac{\sqrt{-t}}{\sqrt{-t + 4M_p^2}}
\end{equation}
where $M_p$ is the proton mass. It means that part of the integral must be evaluated over a domain where the CFF is continued analytically and where experimental measurements are impossible. This does not represent a theoretical issue per se. For instance, the Double Distribution can be characterized from a limited range in $\xi$ as highlighted in \cite{DallOlio:2024vjv}, and then used to construct the CFF in the range where it is not measured. Likewise, the extraction of GPDs on the lattice, which is performed in Euclidean space using a space-like definition of the kinematic variables gives access to any $\xi$ at any $t$ \cite{HadStruc:2024rix}. Yet, in the context of an analysis based on experimental data, the severe kinematic limitation on the information on the CFFs represents a challenge. The flexibility of the neural network parametrization attempts to introduce as little bias as possible in the analytic continuation of the CFFs outside of their experimental determination. 

\begin{figure}
    \centering
    \includegraphics[width=0.99\linewidth]{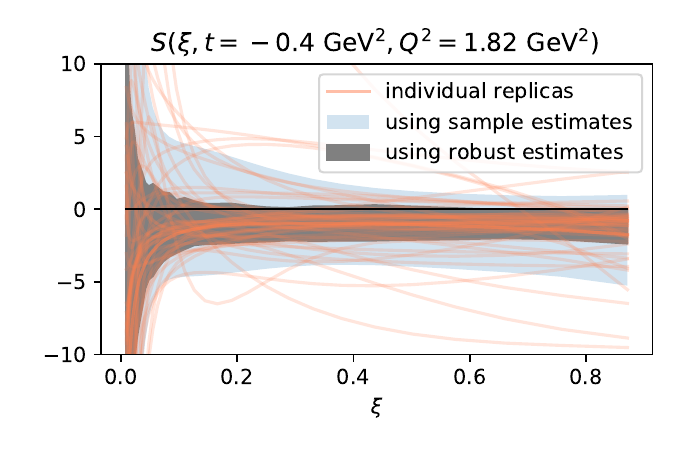}
    \caption{Subtraction constant as a function of $\xi$ at a given value of $(t, Q^2)$ obtained from the dispersion relation applied to a neural network extraction of $\mathcal{H}$ from the world DVCS dataset in 2019. We show a subset of 50 replicas and the uncertainty computed using the ordinary sample standard deviation and the MAD robust estimator.}
    \label{fig:sub}
\end{figure}

From the 100 sets of CFFs stemming from the neural network analysis, we compute 100 samples of the functional distribution of the subtraction constant which we will call \emph{replicas} in the following. The result at one kinematic point $(t, Q^2)$ as a function of $\xi$ is presented in Fig. \ref{fig:sub}. We present the result using both the traditional sample standard deviation estimate:
\begin{equation}
\sigma_{sample} = \frac{1}{\sqrt{N-1}} \sqrt{\sum_{i=1}^{N} (X_i - \textrm{mean}(X))^2}
\end{equation}
and using the outlier robust estimate of the standard deviation known as mean absolute deviation (MAD):
\begin{equation}
    \sigma_{MAD}(X) = \lambda\, \textrm{med}\bigg(|X - \textrm{med}(X)|\bigg)\,.
\end{equation}
med stands for the median, and $\lambda = 1 / \Phi^{-1}(3/4) = 1.4826$ where $\Phi(x)$ is the standard normal cumulative distribution function. This constant rescaling allows the MAD operator to coincide with the standard deviation in infinite statistics under the assumption that the distribution is Gaussian. The large difference between the two estimates in Fig. \ref{fig:sub} highlights the important contamination by outliers, that we elaborate on in the next paragraph. 

\begin{figure}
    \centering
    \includegraphics[width=0.99\linewidth]{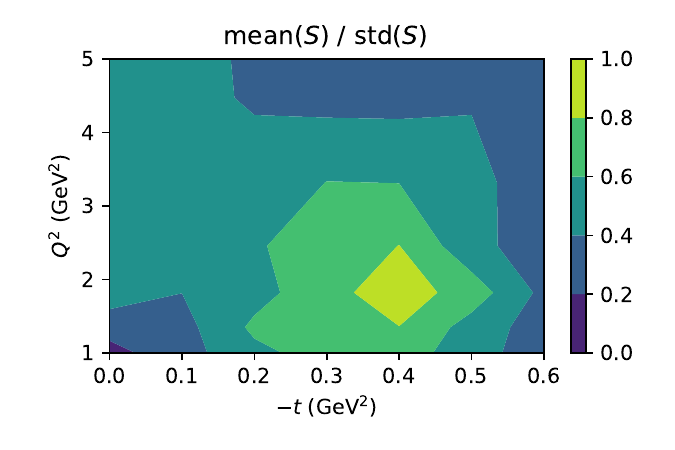}
    \caption{Strength of the signal of the subtraction constant (robust estimates), expressed in number of standard deviations from 0. We use the most precise value of $\xi$ for each kinematic $(t, Q^2)$, excluding $\xi > 0.5$ (see text). Only the best kinematics allow a characterization at $1 \sigma$ from 0.}
    \label{fig:signal}
\end{figure}

Since the real and imaginary parts of the CFFs have been modelled independently without enforcing the connection between them induced by the polynomiality of GPDs, there is in principle no expectation that the subtraction constant will end up independent of $\xi$. However, the result is indeed globally compatible with a constant. We exclude from the analysis the subtraction constant for $\xi > 0.5$, as there seems to be a slight systematic downward shift of the subtraction constant at large $\xi$. This may stem from the fact that $\textrm{Im}\,\mathcal{H}(\xi)$ is not constrained to go to 0 as $\xi \rightarrow 1$, leading to slightly less behaved subtraction constant integrals in this limit. At the value of $(t, Q^2)$ presented in Fig. \ref{fig:sub} -- one of the most precise since it corresponds to a region well explored by the JLab 6 GeV data -- the subtraction constant is still only one standard deviation away from 0 evaluated using outlier robust statistics. We present in Fig. \ref{fig:signal} the strength of the signal of the subtraction constant on the kinematic domain. The fact that neural network analyses of the DVCS dataset available before the JLab 12 GeV upgrade lead to subtraction constants which are compatible with 0 was also established in Ref. \cite{Kumericki:2019ddg}. This results to a large extent from the poor constraints on the real part of the CFF $\mathcal{H}$ within the current experimental dataset. This highlights the interest of a better determination of this quantity, achievable for instance by measuring unpolarised beam charge asymmetry observables with a positron beam at JLab \cite{Accardi:2020swt,Dutrieux:2021ehx}.

\subsection{Treatment of outliers}

As we have already noticed, our data suffers from noticeable outliers. Therefore, instead of the sample means, variances and covariances, we should use outlier robust estimates. In all this analysis, we will use the sample median instead of the sample mean and the MAD instead of the sample standard deviation.

\begin{figure}
    \centering
\includegraphics[width=0.99\linewidth]{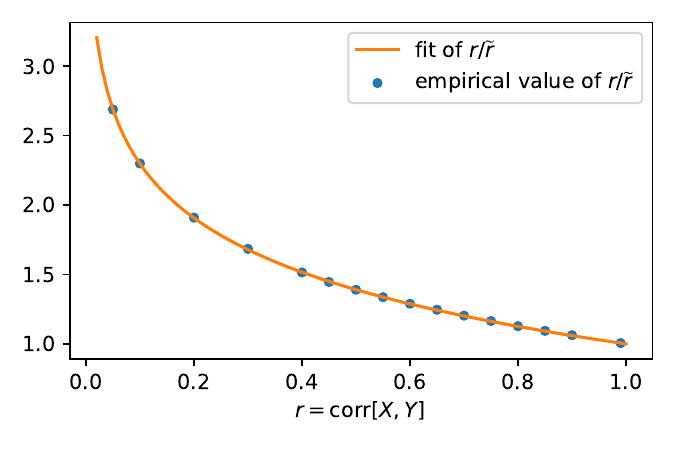}
\caption{Empirical measurements of $r / \widetilde{r}$ from extensive samplings of correlated normal distributions, and their fit by the functional form of eq. \eqref{eq:muR}.}
    \label{fig:fitmu}
\end{figure}

There exists a considerable literature devoted to the question of a robust estimate of the covariance / correlation matrix (see among the most popular suggestions \cite{GK,Rousseeuw,LEDOIT}). We will use in this paper the straightforward generalization of the MAD estimator for the correlation, which we have not seen used before in the existing literature:
\begin{equation}
\widetilde{r} \equiv \frac{\textrm{med}\bigg((X-\textrm{med}(X))(Y-\textrm{med}(Y))\bigg)}{\sqrt{\textrm{med}\bigg((X - \textrm{med}(X))^2\bigg)\textrm{med}\bigg((Y - \textrm{med}(Y))^2\bigg)}} \label{eq:rtilda}
\end{equation}
A comparison of this estimator to the many others presented in the literature would divert us from the physics purpose of this paper, and is conducted in a separate more statistically focused paper \cite{Dutfuture}. We stick therefore to a succint presentation. In effect, this estimator does not directly relate to the correlation $r \equiv \textrm{corr}[X,Y] = \textrm{cov}[X,Y] / (\sigma_X \sigma_Y)$: just like a factor $\lambda$ was necessary to relate the MAD estimator to the standard deviation for a normal distribution, some procedure is needed to match \eqref{eq:rtilda} to $r = \textrm{corr}[X,Y]$. It is likely that there exist no closed-form matching formula in the case of a bivariate normal distribution. However as presented in \cite{Dutfuture} and demonstrated empirically in Fig. \ref{fig:fitmu}, the following approximation is accurate beyond the per mille level and fully satisfactory considering the precision of this study:
\begin{equation}
r / \widetilde{r} = 1 - 0.5635 \ln |r|\label{eq:muR}
\end{equation}
Isolating properly the correlation coefficient yields:
\begin{equation}
r = 0.5635 \, \widetilde{r}\, W\left(\frac{10.47}{ |\widetilde{r}|}\right)
\end{equation}
where $W$ is the Lambert-$W$ function, defined as the inverse function of $f(W) = W e^W$.

One should note that this definition does not guarantee that the resulting covariance / correlation matrix is positive definite. Only in the limit of infinite statistics of a true Gaussian distribution do we formally expect that this property is fulfilled. This may however not be as much of a drawback that it appears at first sight. The maximal size of negative eigenvalues in the spectrum provides a clear physical evidence of the magnitude of poorly estimated correlations. Instead of using the inverse of the covariance matrix to compute the $\chi^2$, one could therefore use a singular value decomposition (SVD) with a threshold on the singular values larger than the absolute value of the negative eigenvalues.

\begin{figure}
    \centering
\includegraphics[width=0.99\linewidth]{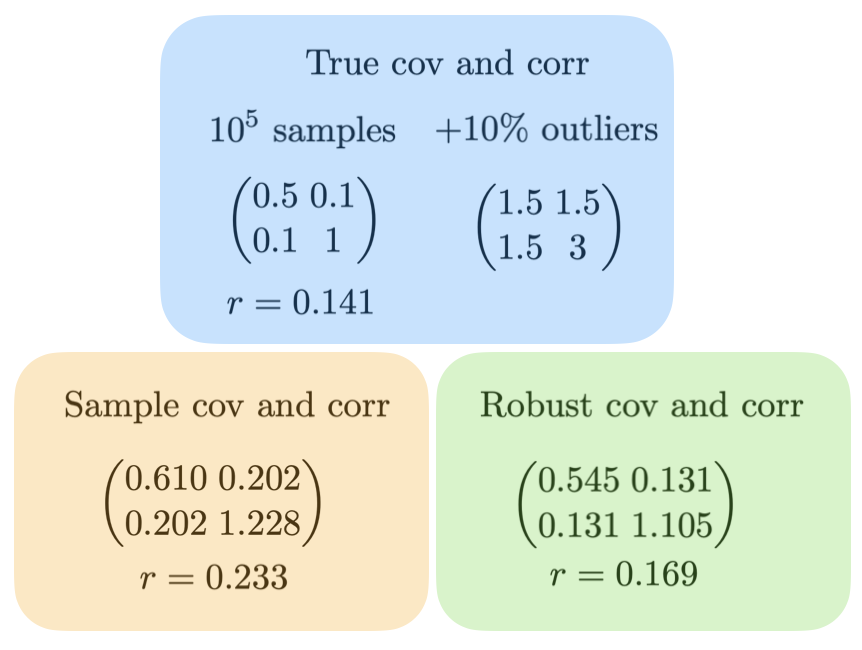}
\caption{We generate a high-statistics sample with 10\% of outliers. The median based estimators are less contaminated by the outliers than the ordinary sample estimators.}
    \label{fig:example_outlier}
\end{figure}

To demonstrate practically the interest of our outlier resilient covariance matrix estimate, we deliberately create an outlier-ridden distribution, made of a mixture of a narrow normal distribution, and a fraction of samples from a wider normal distribution. The result is depicted in Fig. \ref{fig:example_outlier} and shows that our robust estimators are less affected by the outliers. Note that in general, our robust estimators have a larger dispersion than the sample ones (in statistical terms, they are less efficient). In the example presented here, the dispersion of our robust operators is typically 20 to 30\% larger than the sample ones. But although the sample operators are less dispersed, their expectation is further from the value of interest in presence of strong outliers.

\subsection{A model of the $D$-term}

As we have already noted, our data-driven neural network extraction of the subtraction constant is largely unconstrained as $|t|$ and $Q^2$ increase. We have stressed in Section \ref{sec:pressure-distributions-nlo} that the scale dependence is instrumental to perform a model independent extraction of the $D$-term or the $C(t)$ GFF from the experimental data. To proceed forward and obtain sensible results at large $|t|$, we are forced to use a model of $D$-term. We choose the following usual strategy:
\begin{itemize}
\item The scale dependence of the $D$-term is given by the LO renormalization group equation resummed at leading-logarithmic accuracy. We use the LO running of $\alpha_s$ with $\alpha_s(M_Z^2) = 0.118$ and threshold crossing at $m_c = 1.27$ GeV and $m_b = 4.18$ GeV.
\item We truncate the Gegenbauer expansion \eqref{eq:Gegen} to either $n = 1$ or $n = 3$, which allows us to probe a part of the shadow $D$-term uncertainty as we have explained in the previous section.
\item We assume an equal contribution of the light quarks $d_n^{u} = d_n^{d} = d_n^{s} = d_n^{uds}$ and a purely radiatively generated charm contribution, that is $d_n^c(\mu^2 = m_c^2) = 0$.
\item We enforce a factorized $t$-dependence under the form of a tripole Ansatz:
\begin{equation}
    D^a(t) = D^a(t = 0) \left(1-\frac{t}{M^2}\right)^{-3}\,,
\end{equation}
where $M = 0.8$ GeV. The lack of distinctive $t$-dependence in the CFF extraction makes this fully constrained Ansatz satisfactory.
\end{itemize}

Our model is therefore entirely defined by the coefficients $d_n^{uds}(t = 0, \mu_0^2)$ and $d_n^g(t = 0, \mu_0^2)$ at a conventionally fixed scale $\mu_0 = 2$ GeV. We want to obtain the distribution of those parameters so that 
\begin{equation}
\int_{-1}^1 \ud\omega\,T^a(\omega, \alpha_s(Q^2)) \Sigma^{ab}(Q^2, \mu_0^2) \otimes D^b(\omega, t, \mu_0^2)
\end{equation}
approximates as much as possible the distribution of the 100 replicas $S(\xi, t, Q^2)$. $\Sigma^{ab}(Q^2, \mu_0^2)$ represent the leading-logarithmic evolution operator of the $D$-term. Implicit summation on the repeated indices is subtended. 

The problem presents itself as finding the best fit of a target function by a parametrized function. A simple way to proceed is to sample the target function and perform a least-squares fit. However, the answer may depend on the choice of kinematics where the sample is performed. In order to perform a correlated fit, we need to select much fewer kinematic values than the number of replicas which are available. With 100 replicas, we will use $N_{kin}$ kinematics selected because they represent the strongest signal of the subtraction constant. Precisely, we decompose the $(\xi, t, Q^2)$ phase-space in a regular grid with a logarithmic spacing in $\xi$ characterized by a multiplicative factor of 1.5, a uniform spacing in $t$ characterized by a pace of $0.1$ GeV$^2$, and a logarithmic spacing in $Q^2$ (multiplicative factor of 1.35). Then we select the $N_{kin}$ kinematics where the ratio of the median of the replicas by the MAD is the largest. This prevents from overfitting our model on a region where the neural network is left largely unconstrained.

Let's call $(\xi_i, t_i, Q^2_i)$ the set of kinematics we have just described, on which the replicas of $S(\xi, t, Q^2)$ are sampled. Once this choice is fixed, several options present themselves to determine the best parameters.
\begin{enumerate}
\item The most natural option is to determine the distribution of our free parameters so as to minimize the correlated least squares:
\begin{align}
&\sum_{i, i'} (\textrm{model}(\xi_i, t_i, Q^2_i) - \bar{S}(\xi_i, t_i, Q^2_i)) \times  \textrm{cov}^{-1} [S]_{i,i'}  \nonumber \\
& \hspace{40pt}\times (\textrm{model}(\xi_{i'}, t_{i'}, Q^2_{i'}) - \bar{S}(\xi_{i'}, t_{i'}, Q^2_{i'}))\,.
\end{align}
$\bar{S}$ is the sample median of the dataset and $\textrm{cov}[S]$ the robust covariance matrix. Since the fit is linear, the best-fit parameters are normally distributed and we do not need to use individual replicas.
\item If we used an uncorrelated least-squares fit (only the diagonal terms of the covariance matrix), then we would not be limited in the number of kinematics where to perform the fit. However, neglecting outright the statistical information of correlation of the target function seems unjustified.
\item The LO study in Ref. \cite{Dutrieux:2021nlz} used an hybrid approach: for each replica $S_j(\xi, t, Q^2)$ where $1 \leq j \leq 100$ labels the replica, the best-fit value was found with an uncorrelated least squares: \begin{equation}
\sum_i \frac{(\textrm{model}(\xi_i, t_i, Q^2_i) - S_j(\xi_i, t_i, Q^2_i))^2}{(\Delta S(\xi_i, t_i, Q^2_i))^2}   \label{eq:hybrid}    
\end{equation}
where $\Delta S(\xi_i, t_i, Q^2_i)$ is the outlier robust standard deviation computed from the 100 replicas. The result is then made of the distribution of the best-fit value over each replica. This strategy can in principle be applied to an arbitrary number of kinematics, and yet encompasses part of the correlated information through the use of the distribution of replicas. In the absence of outliers distorting the distribution, if we used $\textrm{cov}^{-1}[S]$ instead of $1 / (\Delta S)^2$ in eq. \eqref{eq:hybrid}, we would find exactly the first method, while if we used $\bar{S}$ instead of the individual $S_j$ replicas, we would recover exactly the second method.
\end{enumerate}

\begin{figure}
    \centering
\includegraphics[width=0.99\linewidth]{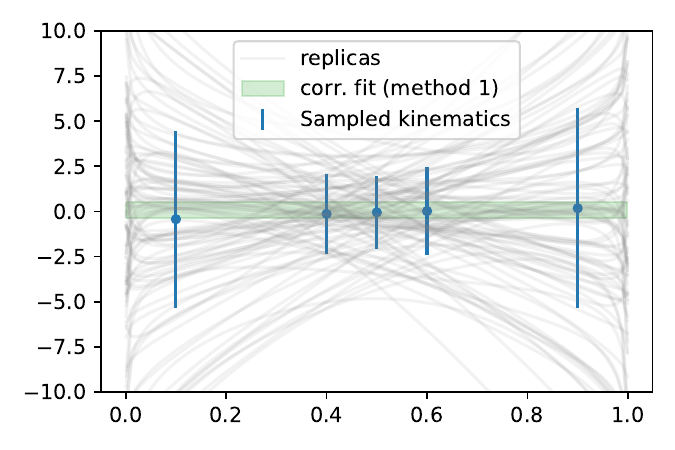}
\includegraphics[width=0.99\linewidth]{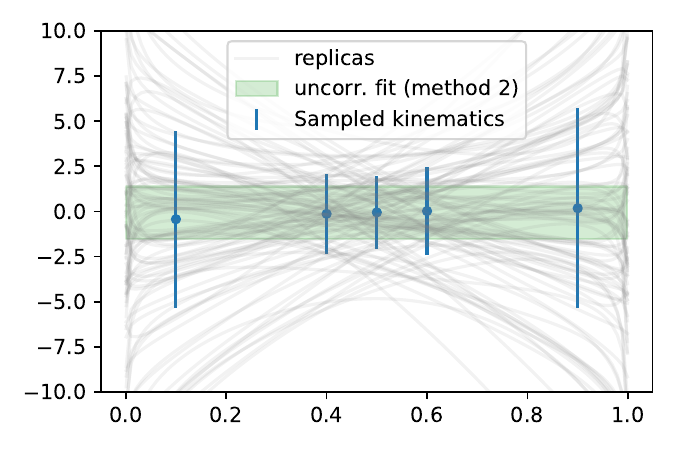}
\includegraphics[width=0.99\linewidth]{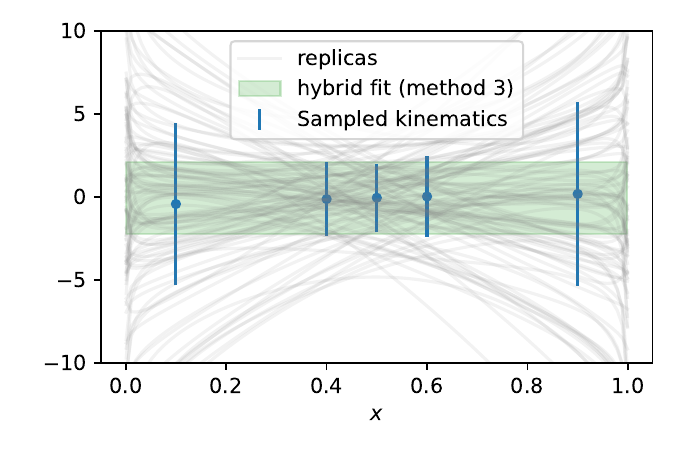}
    \caption{Fit by a constant function of a functional distribution represented by the grey replicas. The fit is performed on the sampled kinematics represented by the blue points using 1) a fully correlated fit, 2) a fully uncorrelated fit and 3) the hybrid method of uncorrelated fit replica-by-replica.}
    \label{fig:methods}
\end{figure}

To appreciate the difference between the three methods independently from the question of the outlier suppression, we construct in Fig. \ref{fig:methods} a fictitious normal distribution of replicas (grey curves), sample it on some kinematics (blue points), and apply the different methods in the absence of outliers. The correlated fit (top plot) exhibits a smaller variance than the uncorrelated one (middle plot) as a reflection of the fact that the grey replicas exhibit a long-distance anti-correlation: the replicas that go down at small $x$ tend to go up at large $x$ and vice-versa. This has the effect of pinning down the best constant more precisely than when this information is simply neglected. The hybrid method (bottom plot) exhibits a significantly larger uncertainty than the other two. In the following, we will present results using the correlated method. It is clear that the most reliable strategy would be to perform a full refit of the experimental data -- which is outside of the scope of this study which only aims at giving a qualitative understanding of the effect of switching from a LO to a NLO analysis.

\begin{figure}
    \centering
\includegraphics[width=0.99\linewidth]{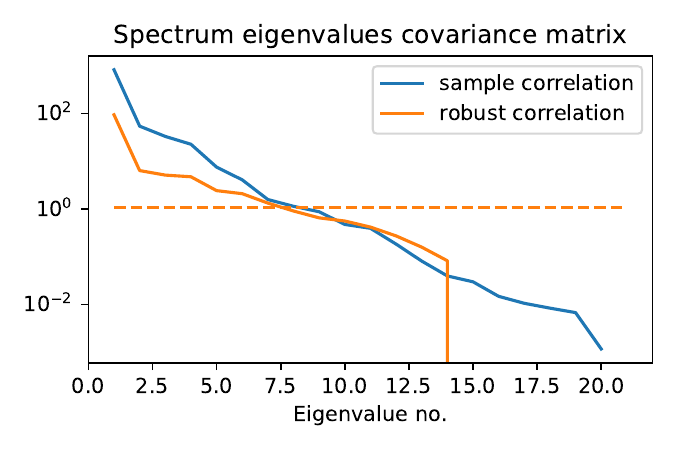}
\caption{Comparison of the spectrum of eigenvalues of the covariance matrix for $N_{kin} = 20$ on the subtration constant dataset on which this study is performed. We compare the sample covariance to our robust estimator. The dotted line represents the largest negative eigenvalue in the spectrum of the robust estimate.}
    \label{fig:compCorr}
\end{figure}

Finally, to stress once again the importance of our robust estimate of the covariance, we plot in Fig. \ref{fig:compCorr} a comparison between the spectrum of eigenvalues of the sample covariance matrix versus our robust estimate with $N_{kin} = 20$ on the subtraction constant dataset. The difference between the largest eigenvalues is mainly driven by the fact that $\sigma_{MAD}$ is smaller than the sample standard deviation which is inflated by outliers. The eigenvalues of the robust estimator then decrease more slowly than the sample one, which means that they will produce an increased stability in a $\chi^2$ fit. Six eigenvalues of the robust estimator are negative. The dotted line represents the largest of them in absolute value, and represents a physical criterion to discard smaller eigenvalues as unreliable. In the end, at most 7 eigenvalues of the covariance matrix are reliably estimated. 

The fits presented in the following section are stable when $N_{kin}$ varies in the interval 10 to 30. Below, we under-sample the phase-space available to the study, resulting in larger uncertainties. Above, the quality of inference of the covariance matrix decreases sharply since $N_{kin}$ becomes fairly large compared to the 100 replicas. When we use the unreliable sample covariance estimator, we do not find such an extended region of stability. We will show all results of the fits for $N_{kin} = 20$ and the robust estimators. $N_{kin} = 20$ corresponds to probing the subtraction constant in the region: $\xi \in [0.1, 0.4]$, $-t \in [0.2, 0.4]$ GeV$^2$ and $Q^2 \in [1, 2.5]$ GeV$^2$, which corresponds to the bulk of the most constraining dataset, JLab 6 GeV.

\subsection{Results of the fits at LO}

\subsubsection{LO radiative gluons and $n = 1$}

At first, we consider only terms with $n = 1$ in the Gegenbauer expansion of eq. \eqref{eq:Gegen}. Furthermore, we assume a radiative gluon generation, that is that $d_1^g(t = 0, \mu_g^2) = 0$ for some low-lying scale $\mu_g^2$. Therefore, $d_1^{uds}(t = 0, \mu_0^2)$ is really the only free parameter. Using $\mu_g = 300$ MeV, we obtain the LO result:
\begin{align}
&d_1^{uds}(t = 0, 2 \textrm{ GeV}^2) = & -0.6 \pm 1.1 \\ \hline
&d_1^g(t = 0, 2 \textrm{ GeV}^2) = & -0.8 \pm 1.5  \nonumber\\
&d_1^c(t = 0, 2 \textrm{ GeV}^2) = & -0.003 \pm 0.005  \nonumber
\tag{LO n=1 radiative gluons}
\end{align}

In spite of the different fitting methodology compared to Ref. \cite{Dutrieux:2021nlz}, the results are very similar: there was determined $d_1^{uds} = -0.5 \pm 1.2$ and $d_1^g = -0.6 \pm 1.6$. It was also noticed in Ref. \cite{Dutrieux:2021nlz} that the threshold $\mu_g$ were gluons are introduced has barely any impact on the fitted value of $d_1^{uds}$. For instance, if we use $\mu_g = 1$ GeV, we still obtain $d_1^{uds}(t = 0, 2\textrm{ GeV}^2) = -0.6 \pm 1.1$, while on the other hand, $d_1^g = -0.1 \pm 0.2$.

In order to understand this interesting observation, we need to remind ourselves that at LO, there is no direct contribution of the gluons to the subtraction constant. The only contribution is indirect, through the radiation of quarks by gluons in the perturbative evolution. On the range of $Q^2$ relevant for this analysis, that is $Q^2 \in [1, 2.5]$ GeV$^2$, the evolution operator resummed to leading logarithmic accuracy reads:
\begin{equation}
\begin{pmatrix}
d_1^{uds}(2.5 \textrm{ GeV}^2) \\ d_1^{g}(2.5 \textrm{ GeV}^2) \\ d_1^{c}(2.5 \textrm{ GeV}^2)
\end{pmatrix} = \begin{pmatrix}
0.92 & 0.015 \\
0.23 & 0.95 \\
0.001 & 0.007
\end{pmatrix} \begin{pmatrix}
d_1^{uds}(1 \textrm{ GeV}^2) \\ d_1^{g}(1 \textrm{ GeV}^2)
\end{pmatrix}\,.
\end{equation}
We label the coefficients of the evolution matrix from $\mu_0^2$ to $\mu^2$ as:
\begin{equation}
\begin{pmatrix}
\Gamma_1^{qq}(\mu^2, \mu_0^2) & \Gamma_1^{qg}(\mu^2, \mu_0^2) \\
\Gamma_1^{gq}(\mu^2, \mu_0^2) & \Gamma_1^{gg}(\mu^2, \mu_0^2) \\
\Gamma_1^{cq}(\mu^2, \mu_0^2) & \Gamma_1^{cg}(\mu^2, \mu_0^2)
\end{pmatrix}\,.
\end{equation}
Notice how small $\Gamma_1^{qg}$, the radiation of light quarks by gluons, is in the range of scales covered by the bulk of the experimental data. This means that gluon contribution to the subtraction constant at LO is heavily suppressed. Introducing the gluon radiation threshold, we obtain that:
\begin{align}d_1^{uds}(\mu^2) = [\Gamma_1^{qq}(\mu^2, \mu_0^2) +\Gamma_1^{qg}(\mu^2, \mu_0^2)\Gamma_1^{gq}(\mu_0^2, \mu_g^2) \nonumber \\ / \Gamma_1^{qq}(\mu_0^2, \mu_g^2)] \times  d_1^{uds}(\mu_0^2)\,.\end{align}
The maximal effect of gluons on the fit of $d_1^{uds}$ is obtained when $\mu_0^2$ and $\mu^2$ are taken at the extreme values covered reliably by the experimental data, so here 1 and 2.5 GeV$^2$. We find therefore that the features of the fitted data in the interval $[1, 2.5]$ GeV$^2$ that can be imputable to gluons are typically of the order of $\Gamma_1^{qg}(2.5, 1) / \Gamma_1^{qq}(2.5,1) \times \Gamma_1^{gq}(1, 0.09) / \Gamma_1^{qq}(1, 0.09) = 0.015 / 0.92 \times 1.21 = 2$~\% of the contribution imputable to $d_1^{uds}$. If $\mu_g$ increases, the gluonic contribution decreases even more, but that is in any case completely imperceptible. In other words, in a LO analysis, radiative gluons might as well be equivalent to no gluons at all. Although this means that the quark contribution to the GFF $\sum_q C_q(t)$ is quite independent of the choice of radiative threshold, it clearly means that the overall GFF $C(t) = \sum_q C_q(t) + C_g(t)$ is extremely unreliable.

\subsubsection{LO radiative gluons and $n = 3$}

Still using a radiative gluon generation with $\mu_g = 300$ MeV, we now allow both $d_1^{uds}(t = 0, 2$ GeV$^2)$ and $d_3^{uds}(t = 0, 2$ GeV$^2)$ to be fitted. At LO, we find:
\begin{align}
&d_1^{uds}(t = 0, 2 \textrm{ GeV}^2) = & -2.1 \pm 26.6 \nonumber\\
&d_3^{uds}(t = 0, 2 \textrm{ GeV}^2) = & 1.5 \pm 26.5  \\  \hline
&d_1^g(t = 0, 2 \textrm{ GeV}^2) = & -2.9 \pm 37  \nonumber \\&d_3^g(t = 0, 2 \textrm{ GeV}^2) = & 0.2 \pm 4.1  \nonumber 
\tag{LO n=3 radiative gluons}
\end{align}
$d_1$ and $d_3$ are anti-correlated in excess of 99\% as identified in Ref. \cite{Dutrieux:2021nlz} before. One observes that, within uncertainty, $d_1^{uds} \approx - d_3^{uds}$. In other words, the uncertainty in the extraction is almost entirely stemming from contamination of LO shadow $D$-terms. We have derived in Section 4 an approximate estimator of the uncertainty linked precisely to this shadow $D$-term $d_1^{uds} \approx - d_3^{uds}$ with a simplified evolution kernel. We found that \eqref{eq:backenvest}:
\begin{equation}
\sigma_{d1q} \approx \sigma_{d3q} \approx \frac{\Delta S}{\displaystyle \left(1-\frac{\alpha_s(Q_{max}^2)}{\alpha_s(Q^2_{min})}\right)}
\end{equation}
$\Delta S$ can be obtained by noting that the fit with $d_1$ alone reads $S = 8 / 3 \times d_1^{uds}$, and therefore $\Delta S \approx 8 / 3 \times 1.1$. This is probably an underestimation, since it takes the uncertainty of the simplest fit as a measure of the uncertainty of the full quantity. Then using $Q^2_{max} = 2.5$ GeV$^2$ and $Q^2_{min} = 1$ GeV$^2$, we find 
\begin{equation}
\sigma_{d1q} \approx \sigma_{d3q} \approx 16\,,
\end{equation}
to compare with the value of 26.5 that our fit produced. Besides the likely underestimation of $\Delta S$, the main drawback of this approximation is the reduction of the information contained in the scale dependence to a sole interval $[Q^2_{min}, Q^2_{max}]$ where we assume that the data is uncorrelated and uniformly constraining. Then, the result is of course sensitive to the choice of this interval. For instance, simply raising $Q^2_{min}$ to 1.4 GeV$^2$ [and therefore reducing the range in scales where we believe the data to be truly constraining] would produce an estimate of $\sigma_{d1q} \approx \sigma_{d3q} \approx 25.5$, very similar to the one truly observed. The approximate evolution used to derive our estimate \eqref{eq:backenvest} only represents a minor imprecision owing to the negligible effect of radiative gluons and the fact that our scales are close to the charm mass.

\subsubsection{LO free gluons and $n = 1$}

\begin{figure}
    \centering
\includegraphics[width=0.99\linewidth]{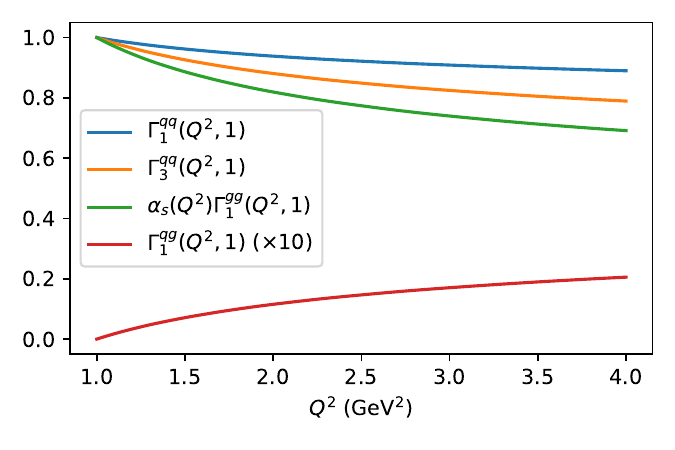}
\caption{Comparison of the functional form of the operators $\Gamma_1^{qq}$, $\Gamma_3^{qq}$, $\alpha_s \Gamma_1^{gg}$ and $\Gamma_1^{qg}$ as a function of the scale. The essence of the deconvolution problem is that, when the fitted functional forms are too similar to one another, the associated parameters are extremely difficult to differentiate and inflate considerably the uncertainty.}
    \label{fig:funcform}
\end{figure}

This time, we allow $d_1^g(t = 0, \mu_0^2)$ to also be a free parameter. We find at LO:
\begin{align}
&d_1^{uds}(t = 0, 2 \textrm{ GeV}^2) = & -0.6 \pm 1.1 \nonumber\\
&d_1^g(t = 0, 2 \textrm{ GeV}^2) = & -11 \pm 132  \\  \hline
&d_1^c(t = 0, 2 \textrm{ GeV}^2) = & -0.04 \pm 0.47  \nonumber
\tag{LO n=1 free gluons}
\end{align}
There again, the results are in good agreement with the results of Ref. \cite{Dutrieux:2021nlz}. Two interesting features are noticeable: the extraction of $d_1^{uds}$ has been left unchanged by the addition of free gluons, and the uncertainty of the gluon term has increased by a factor 90 compared to the radiative gluons. This factor of 90 can be understood as being related to $\Sigma^{qq}(2.5, 1) / \Sigma^{gq}(2.5, 1) = 0.92/0.015 \approx 60$, the factor by which $d_1^g(1$~GeV$^2)$ must be larger than $d_1^{uds}(1$~GeV$^2)$ so that both terms contribute with the same order of magnitude to the fitting of the data. As for the reason why $d_1^{uds}$ is unchanged by free gluons at LO, it comes from the fact that the fitted distributions of $d_1^{uds}$ and $d_1^g$ are largely uncorrelated (correlation coefficient of -0.10). As we have stressed before, the large similarity of $\Gamma_1^{qq}$ and $\Gamma_3^{qq}$ is the root of the very large correlation between $d_1^q$ and $d_3^q$. On the other hand, $\Gamma_1^{qq}$ and $\Gamma_1^{qg}$ present very different functional forms, and result therefore in far less correlated fits as can be observed on Fig. \ref{fig:funcform}.

\subsection{Results of the fits at NLO}

At NLO, for $\mu^2 = Q^2$ and with a truncation up to Gegenbauer moments of order $n = 3$, the subtraction constant reads:
\begin{align}
S = \sum_q e^2_q S^q + S^g\,,  
\end{align}
\begin{align}
S^q \eqNLO d_1^q \left(4 - \frac{4}{9}\frac{\alpha_s C_F}{4 \pi}\right) + d_3^q \left(4 + \frac{14759}{450}\frac{\alpha_s C_F}{4 \pi}\right)\,, \label{eq:NLOq}
\end{align}
\begin{align}
S^g \eqNLO \frac{\sum_q e^2_q \alpha_s T_F}{4\pi} \left(-\frac{172}{9}d_1^g - \frac{3317}{150} d_3^g\right)\,,
\label{eq:NLOg}
\end{align}
where $C_F = 4/3$ and $T_F = 1/2$. We use the LO running of $\alpha_s$ from APFEL \cite{Bertone:2013vaa,Bertone:2017gds}, which remains continuous at the heavy quark mass thresholds. However, a naive implementation of the gluon coefficient function is discontinuous at threshold due to the factor $\sum_q e^2_q$. This makes no practical numerical difference for radiative gluons fits, but becomes important for the fit results with free gluons. The appropriate course of action would be to include heavy quark mass effects in the coefficient function, but this extends beyond the scope of this paper. To avoid spurious effects, we will therefore consider for the rest of the paper that the factor $\sum_q e^2_q$ in the gluonic contribution to the subtraction constant is fixed to 10/9, the value it assumes naively if $n_f = 4$. 

\subsubsection{NLO radiative gluons and $n = 1$}

With a threshold of radiative gluon generation at 300 MeV, we find:
\begin{align}
&d_1^{uds}(t = 0, 2 \textrm{ GeV}^2) = & -0.7 \pm 1.3 \\ \hline
&d_1^g(t = 0, 2 \textrm{ GeV}^2) = & -0.9 \pm 1.8  \nonumber\\
&d_1^c(t = 0, 2 \textrm{ GeV}^2) = & -0.003 \pm 0.006  \nonumber
\tag{NLO n=1 radiative gluons}
\end{align}
The results are almost identical to the LO results of the same fit. There again, changing the threshold for gluon production is only a little effect on $d_1^{uds}$, although larger than at LO. With a threshold $\mu_g^2 = 1$ GeV$^2$, we find $d_1^{uds} = -0.6 \pm 1.1$ and $d_1^g = -0.1 \pm 0.2$. 

To understand the similarity between the LO and NLO fit in the case where only $d_1^{uds}$ is fitted, we observe that $\alpha_s$ is at most $\alpha_{s, max} = 0.35$ in the fitted range. Then the NLO quark term reads \eqref{eq:NLOq}:
\begin{equation}
\sum_q e^2_q\left(4 - \frac{4}{9}\frac{\alpha_{s, max} C_F}{4 \pi}\right) \approx 3.98 \sum_q e^2_q \approx 2.65\,,
\end{equation}
where we neglected the charm contribution to the subtraction constant, and for gluons \eqref{eq:NLOg}:
\begin{align}
-\frac{172}{9}\frac{\sum_q e^2_q \alpha_{s, max} T_F}{4\pi} &\approx -0.30\,,
\end{align}
which is of the order of 10\% of the quark contribution. Due to its negative sign, it causes a slight increase in the fitted value of $d_1^{uds}$ compared to the situation at LO when the gluon threshold is small enough. Let us note that although gluons still play a minor role in the extraction, it is a much bigger one than at LO where we estimated it to 2\% because of the smallness of $\Gamma_1^{qg}$.

\subsubsection{NLO radiative gluons and $n = 3$}

Still with a radiative gluon threshold at $\mu_g = 300$~MeV, and allowing $d_1^{uds}$ and $d_3^{uds}$ to be fitted, we obtain:
\begin{align}
&d_1^{uds}(t = 0, 2 \textrm{ GeV}^2) = & -1.7 \pm 21 \nonumber\\
&d_3^{uds}(t = 0, 2 \textrm{ GeV}^2) = & 0.7 \pm 15  \\  \hline
&d_1^g(t = 0, 2 \textrm{ GeV}^2) = & -2 \pm 30  \nonumber \\&d_3^g(t = 0, 2 \textrm{ GeV}^2) = & 0.1 \pm 2.3  \nonumber 
\tag{NLO n=3 radiative gluons}
\end{align}
The general uncertainty is of the same order of magnitude as the one obtained at LO and the anti-correlation still in excess of 99\%. However, whereas at LO, we had $d_1^{uds} \approx - d_3^{uds}$, the situation has changed a bit. Since the subtraction constant has been modified at NLO, it does not admit exactly the same shadow $D$-terms. Indeed, the $d_3^{uds}$ term in \eqref{eq:NLOq} gives with $\alpha_s(2 \textrm{ GeV}^2) \approx 0.3$:
\begin{equation}
\sum_q e^2_q\left(4 + \frac{14759}{450}\frac{\alpha_s C_F}{4 \pi}\right) \approx 3.36\,.
\end{equation}
An NLO shadow $D$-term must now also cancel the contribution stemming from the gluons, which we can no longer ignore even in the radiative approximation. Using the reference scale of 2 GeV$^2$ and the explicit relation between $d_n^g$ and $d_n^{uds}$ offered by the radiation threshold, the gluonic contribution of $n = 1$ reads:
\begin{align}
&-\frac{172}{9}\frac{\sum_q e^2_q \alpha_s T_F}{4\pi} \frac{\Gamma_1^{gq}(2, 0.09)}{\Gamma_1^{qq}(2, 0.09)} d_1^{uds} \approx -0.36 d_1^{uds}\,,
\end{align}
whereas for $n = 3$:
\begin{align}
&\frac{3317}{150}\frac{\sum_q e^2_q \alpha_s T_F}{4\pi} \frac{\Gamma_3^{gq}(2, 0.09)}{\Gamma_3^{qq}(2, 0.09)} d_3^{uds}\approx 0.05 d_3^{uds}\,,\end{align}
We note in passing that the effect of gluons at NLO on $d_3^{uds}$ is much smaller than on $d_1^{uds}$. Finally, the NLO subtraction constant at 2 GeV$^2$ reads approximately as: 
\begin{equation}
S(2 \textrm{ GeV}^2) \eqNLO  (2.65 - 0.36) d_1^{uds} + (3.36 + 0.05) d_3^{uds}\,,
\end{equation}
to compare with the LO:
\begin{equation}
S \eqLO  2.65 d_1^{uds} + 2.65 d_3^{uds}\,.
\end{equation}
An expectation of NLO shadow $D$-term with radiative gluons at a 300 MeV threshold is therefore:
\begin{equation}
d_1^{uds} \approx -\frac{3.36 + 0.05}{2.65 - 0.36} d_3^{uds} \approx -1.5 d_3^{uds}\,.
\end{equation}
Using the observed value of $\sigma_{d3q} = 15$, this would predict $\sigma_{d1q} \approx 22.5$, which is close to the true value of 21. The estimator is made somewhat more complicated than at LO by the necessary inclusion of the gluon term and $\alpha_s$. It provides however reliable results, demonstrating that the interpretation of the uncertainty in terms of a simple shadow $D$-term is a valuable tool.

\subsubsection{NLO free gluons and $n = 1$}

If we now allow the $d_1^g$ term to be freely fitted alongside $d_1^{uds}$, we find:
\begin{align}
&d_1^{uds}(t = 0, 2 \textrm{ GeV}^2) = & -1.1 \pm 7.7 \nonumber\\
&d_1^g(t = 0, 2 \textrm{ GeV}^2) = & -6 \pm 78  \\  \hline
&d_1^c(t = 0, 2 \textrm{ GeV}^2) = & -0.02 \pm 0.27 \nonumber
\tag{NLO n=1 free gluons}
\end{align}
One will notice that the uncertainty on $d_1^{uds}$ has increased by a large factor compared to the case where $d_1^g$ was not a free parameter. This indicates very large correlation between $d_1^{uds}$ and $d_1^g$ at NLO, and the impact of an underlying shadow $D$-term.

The reason why a shadow $D$-term produces a large effect at NLO in the joint fit of $d_1^{uds}$ and $d_1^g$ whereas it was not visible at LO is that gluons now contribute to the subtraction constant in their own right, mostly through $\alpha_s(Q^2)\Gamma^{gg}_1(Q^2, \mu_0^2)$. At LO they could only contribute through the radiation term $\Gamma^{qg}_1(Q^2, \mu_0^2)$. But $\Gamma^{gg}_1$ is a diagonal term in the evolution matrix, whose functional dependence is very similar to that of $\Gamma_1^{qq}$, and fundamentally different of the off-diagonal term $\Gamma_1^{qg}$. The operators can be compared on Fig. \ref{fig:funcform} where it will be apparent that the impact of the shadow $D$-term related to $d_1^g$ at NLO remains smaller than that of $d_3^{uds}$. Therefore $\sigma_{d1q}$ remains less affected by the inclusion of a free NLO $d_1^q$ than by a LO or NLO $d_3^{uds}$.


%% file: section6.tex
In Section \ref{sec:pressure-distributions-nlo}, we derived a simple estimate of the uncertainty of $d_1^q$ and $d_3^q$ when they are fitted jointly in a LO framework with simplified evolution over a range $[Q^2_{min}, Q^2_{max}]$. As we studied NLO fits in Section \eqref{sec:pressure-shadow}, we extended the concept and started to consider the case of explicit gluonic degrees of freedom. Let us give here final general expressions and apply them on a kinematic range relevant for the EIC.

We assume that the contribution of heavy quarks remains always negligible in the subtraction constant and that the contribution of all three light flavors is the same $d_n^{uds}$. In the absence of appropriate heavy quark mass effects in the gluon coefficient function, we fix $n_e = \sum_q e^2_q = 10/9$. Then, at a given value of $t$, we remind that the subtraction constant at NLO truncated to the Gegenbauer moments $n = 3$ reads:
\begin{align}
&S \eqNLO \frac{2}{3} d_1^{uds} \left(4 - \frac{4}{9}\frac{\alpha_s C_F}{4 \pi}\right) -\frac{172}{9}\frac{n_e \alpha_s T_F}{4\pi} d_1^g \nonumber \\
&+ \frac{2}{3} d_3^{uds} \left(4 + \frac{14759}{450}\frac{\alpha_s C_F}{4 \pi}\right) - \frac{3317}{150}
\frac{n_e \alpha_s T_F}{4\pi} d_3^g
\label{eq:er}
\end{align}
where $S$, $d_n^a$ and $\alpha_s$ have all an implicit dependence on $Q^2$. 

\begin{figure}
    \centering
\includegraphics[width=0.99\linewidth]{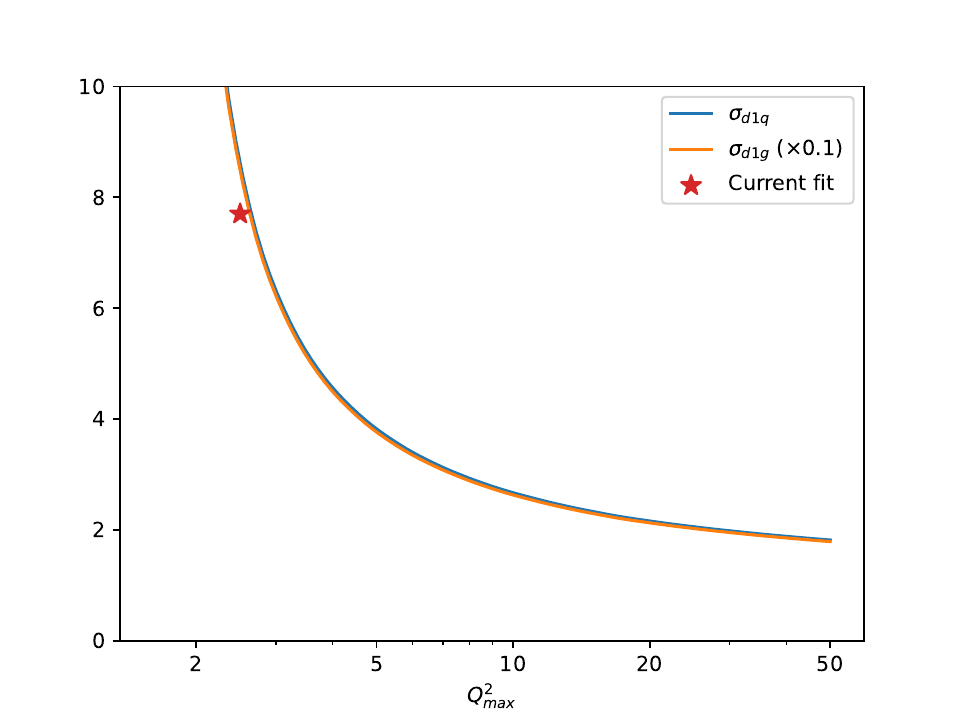}
\caption{Evolution of the uncertainty of $d_1^{uds}$ and $d^1_g$ when the latter is a free parameter depending on the range of $Q^2$ available for the extraction of CFFs. The red star denotes approximately the current situation.}
    \label{fig:gluonunc}
\end{figure}

First let us study the impact of EIC kinematics on the free extraction of a gluon contribution with $n=1$ only. Following the reasoning of Section \ref{sec:pressure-distributions-nlo}, we cancel the subtraction constant at some reference scale:
\begin{align}
0 = a(\mu_0^2) d_1^{uds}(\mu_0^2) + b(\mu_0^2) d_1^g(\mu_0^2)\,,
\end{align}
where the coefficients $a$ and $b$ are read straightforwardly from the first line of eq. \eqref{eq:er}. We choose $\mu_0^2 = m_c^2$, which will also serve as the $Q^2_{min}$ of our data. Then at any scale, we have:
\begin{align}
&S(Q^2) = a(Q^2) (\Gamma_1^{qq}(Q^2, \mu_0^2)d_1^{uds}(\mu_0^2)\nonumber \\ &+ \Gamma_1^{qg}(Q^2, \mu_0^2) d_1^{g}(\mu_0^2)) + b(Q^2) (\Gamma_1^{gq}(Q^2, \mu_0^2)d_1^{uds}(\mu_0^2) \nonumber \\ &+ \Gamma_1^{gg}(Q^2, \mu_0^2)d_1^{g}(\mu_0^2))\,.
\end{align}
Assuming that a quantity $\Delta S$ represents the typical experimental uncertainty on the subtraction constant for $Q^2 \in [\mu_0^2, Q^2_{max}]$ where $Q^2_{max}$ is the largest scale where the subtraction constant is extracted reliably, we find:
\begin{align}
&\sigma_{d1q}(\mu_0^2) = \Delta S  \times \bigg| a(Q_{max}^2) \Gamma_1^{qq} - \frac{a(Q_{max}^2)a(\mu_0^2)}{b(\mu_0^2)} \Gamma_1^{qg} \nonumber \\ &+ b(Q_{max}^2) \Gamma_1^{gq} - \frac{a(\mu_0^2)b(Q_{max}^2)}{b(\mu_0^2)} \Gamma_1^{gg}\bigg| ^{-1}\,,
\end{align}
where all the $\Gamma_n$ operators have an implicit argument $(Q_{max}^2, \mu_0^2)$. Straightforwardly,
\begin{equation}
\sigma_{d1g}(\mu_0^2) = \left| \frac{a(\mu_0^2)}{b(\mu_0^2)}\right| \sigma_{d1q}(\mu_0^2) \,.
\end{equation}
We depict in Fig. \ref{fig:gluonunc} the value of this estimate of the precision using $\Delta S = 3$, that is considering that the current precision on the subtraction constant is extended to a much larger range in $Q^2$. The current estimated precision with the CFF extraction used in this study is depicted by the red star. One observes that constraining the CFFs up to $Q^2 = 20$ Gev$^2$ could reduce the uncertainty by a factor 3 to 4. Let us notice also that our estimator predicts very accurately that $\sigma_{d1g} \approx 10\sigma_{d1q}$, a relation that one can observe to be accurately verified in our actual fit as well.

\begin{figure}
    \centering
\includegraphics[width=0.99\linewidth]{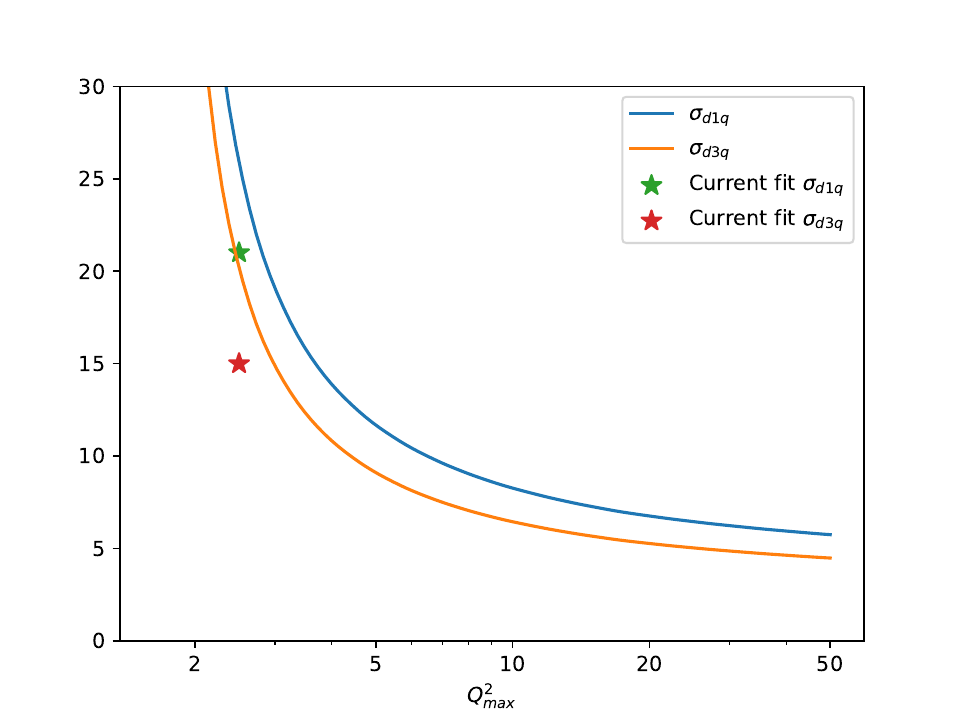}
\caption{Evolution of the uncertainty of $d_1^{uds}$ and $d_3^{uds}$ depending on the range of $Q^2$ available for the extraction of CFFs, considering only the role of shadow $D$-terms with no explicit gluon contribution. The red and green stars denote approximately the current situation obtained when fitting with radiative gluons.}
    \label{fig:quarkunc}
\end{figure}

Now we study the impact of EIC kinematics on the extraction of both $d_1^{uds}$ and $d_3^{uds}$. We will assume that the gluon part of the shadow $D$-term is 0 at the reference scale. Then:
\begin{align}
0 = a(\mu_0^2) d_1^{uds}(\mu_0^2) + c(\mu_0^2) d_3^{uds}(\mu_0^2)\,,
\end{align}
and
\begin{align}
&S(Q^2) = [a(Q^2)\Gamma_1^{qq}(Q^2, \mu_0^2) +b(Q^2)\Gamma_1^{gq}(Q^2, \mu_0^2)] \nonumber \\ &\times d_1^{uds}(\mu_0^2) + [c(Q^2)\Gamma_3^{qq}(Q^2, \mu_0^2) +d(Q^2)\Gamma_3^{gq}(Q^2, \mu_0^2)] \nonumber \\
&\times d_3^{uds}(\mu_0^2)\,.
\end{align}
Combining the two expressions gives the estimator:
\begin{align}
&\sigma_{d1q}(\mu_0^2) = \Delta S \times \bigg|a(Q^2_{max})\Gamma_1^{qq} + b(Q^2_{max}) \Gamma_1^{gq} \nonumber \\
&  - \frac{c(Q^2_{max})a(\mu_0^2)}{c(\mu_0^2)} \Gamma_3^{qq} - \frac{d(Q^2_{max})a(\mu_0^2)}{c(\mu_0^2)} \Gamma_3^{gq} \bigg|^{-1}
\end{align}
We produce the corresponding plot in Fig. \ref{fig:quarkunc}. Constraining the CFFs up to $Q^2 = 20$ GeV$^2$ would now result in a decrease of uncertainty by a factor 2 to 3. We remind that we have only considered here shadow $D$-terms with no explicit gluon contribution, which corresponds to the uncertainty of the fits we have performed earlier with radiative gluons. Adding freedom of explicit gluon contributions would result in yet a far larger increase of uncertainty, combining the effect of Figs. \ref{fig:gluonunc} and \ref{fig:quarkunc}.

However, let us stress that our estimator is only focusing on the impact of the measured range of scales $Q^2$. The EIC will also bring precious high-quality data in regions in $\xi$ which are poorly constrained so far, which will likely decrease the value of $\Delta S$. More work to estimate this impact remains to be done.


%% file: subtraction-nlo.bbl
\begin{thebibliography}{10}

\bibitem{Rutherford:1919fnt}
E.~Rutherford.
\newblock {Collision of \ensuremath{\alpha} particles with light atoms. IV. An
  anomalous effect in nitrogen}.
\newblock {\em Phil. Mag. Ser. 6}, 37:581--587, 1919.

\bibitem{Polyakov:2018zvc}
Maxim~V. Polyakov and Peter Schweitzer.
\newblock {Forces inside hadrons: pressure, surface tension, mechanical radius,
  and all that}.
\newblock {\em Int. J. Mod. Phys. A}, 33(26):1830025, 2018.

\bibitem{Lorce:2018egm}
C\'edric Lorc\'e, Herv\'e Moutarde, and Arkadiusz~P. Trawi\'nski.
\newblock {Revisiting the mechanical properties of the nucleon}.
\newblock {\em Eur. Phys. J. C}, 79(1):89, 2019.

\bibitem{Freese:2021jqs}
Adam Freese.
\newblock {Noether\textquoteright{}s theorems and the energy-momentum tensor in
  quantum gauge theories}.
\newblock {\em Phys. Rev. D}, 106(12):125012, 2022.

\bibitem{Burkert:2018bqq}
V.~D. Burkert, L.~Elouadrhiri, and F.~X. Girod.
\newblock {The pressure distribution inside the proton}.
\newblock {\em Nature}, 557(7705):396--399, 2018.

\bibitem{Kumericki:2019ddg}
Krešimir Kumerički.
\newblock {Measurability of pressure inside the proton}.
\newblock {\em Nature}, 570(7759):E1--E2, 2019.

\bibitem{Dutrieux:2021nlz}
H.~Dutrieux, C.~Lorc\'e, H.~Moutarde, P.~Sznajder, A.~Trawi\'nski, and
  J.~Wagner.
\newblock {Phenomenological assessment of proton mechanical properties from
  deeply virtual Compton scattering}.
\newblock {\em Eur. Phys. J. C}, 81(4):300, 2021.

\bibitem{Alexandrou:2017oeh}
C.~Alexandrou, M.~Constantinou, K.~Hadjiyiannakou, K.~Jansen, C.~Kallidonis,
  G.~Koutsou, A.~Vaquero Avil\'es-Casco, and C.~Wiese.
\newblock {Nucleon Spin and Momentum Decomposition Using Lattice QCD
  Simulations}.
\newblock {\em Phys. Rev. Lett.}, 119(14):142002, 2017.

\bibitem{Alexandrou:2020sml}
C.~Alexandrou, S.~Bacchio, M.~Constantinou, J.~Finkenrath, K.~Hadjiyiannakou,
  K.~Jansen, G.~Koutsou, H.~Panagopoulos, and G.~Spanoudes.
\newblock {Complete flavor decomposition of the spin and momentum fraction of
  the proton using lattice QCD simulations at physical pion mass}.
\newblock {\em Phys. Rev. D}, 101(9):094513, 2020.

\bibitem{Wang:2021vqy}
Gen Wang, Yi-Bo Yang, Jian Liang, Terrence Draper, and Keh-Fei Liu.
\newblock {Proton momentum and angular momentum decompositions with overlap
  fermions}.
\newblock {\em Phys. Rev. D}, 106(1):014512, 2022.

\bibitem{Nair:2024fit}
Sreeraj Nair, Chandan Mondal, Siqi Xu, Xingbo Zhao, Asmita Mukherjee, and
  James~P. Vary.
\newblock {Gravitational form factors and mechanical properties of quarks in
  protons: A basis light-front quantization approach}.
\newblock {\em Phys. Rev. D}, 110(5):056027, 2024.

\bibitem{Yao:2024ixu}
Z.~Q. Yao, Y.~Z. Xu, D.~Binosi, Z.~F. Cui, M.~Ding, K.~Raya, C.~D. Roberts,
  J.~Rodr\'\i{}guez-Quintero, and S.~M. Schmidt.
\newblock {Nucleon Gravitational Form Factors}.
\newblock 9 2024.

\bibitem{Ji:1996ek}
Xiang-Dong Ji.
\newblock {Gauge-Invariant Decomposition of Nucleon Spin}.
\newblock {\em Phys. Rev. Lett.}, 78:610--613, 1997.

\bibitem{Mueller:1998fv}
Dieter Mueller, D.~Robaschik, B.~Geyer, F.M. Dittes, and J.~Ho\v{r}e\v{j}si.
\newblock {Wave functions, evolution equations and evolution kernels from light
  ray operators of QCD}.
\newblock {\em Fortsch.Phys.}, 42:101--141, 1994.

\bibitem{Ji:1996nm}
Xiang-Dong Ji.
\newblock {Deeply virtual Compton scattering}.
\newblock {\em Phys.Rev.}, D55:7114--7125, 1997.

\bibitem{Radyushkin:1996ru}
A.V. Radyushkin.
\newblock {Asymmetric gluon distributions and hard diffractive
  electroproduction}.
\newblock {\em Phys.Lett.}, B385:333--342, 1996.

\bibitem{Radyushkin:1997ki}
A.V. Radyushkin.
\newblock {Nonforward parton distributions}.
\newblock {\em Phys.Rev.}, D56:5524--5557, 1997.

\bibitem{Collins:1996fb}
John~C. Collins, Leonid Frankfurt, and Mark Strikman.
\newblock {Factorization for hard exclusive electroproduction of mesons in
  QCD}.
\newblock {\em Phys.Rev.}, D56:2982--3006, 1997.

\bibitem{Collins:1998be}
John~C. Collins and Andreas Freund.
\newblock {Proof of factorization for deeply virtual Compton scattering in
  QCD}.
\newblock {\em Phys.Rev.}, D59:074009, 1999.

\bibitem{Berger:2001xd}
Edgar~R. Berger, M.~Diehl, and B.~Pire.
\newblock {Time - like Compton scattering: Exclusive photoproduction of lepton
  pairs}.
\newblock {\em Eur.Phys.J.}, C23:675--689, 2002.

\bibitem{Mueller:2013caa}
Dieter Müller, Tobias Lautenschlager, Kornelija Passek-Kumericki, and Andreas
  Schaefer.
\newblock {Towards a fitting procedure to deeply virtual meson production - the
  next-to-leading order case}.
\newblock {\em Nucl. Phys.}, B884:438--546, 2014.

\bibitem{Boussarie:2016qop}
R.~Boussarie, B.~Pire, L.~Szymanowski, and S.~Wallon.
\newblock {Exclusive photoproduction of a $\gamma\,\rho$ pair with a large
  invariant mass}.
\newblock {\em JHEP}, 02:054, 2017.
\newblock [Erratum: JHEP 10, 029 (2018)].

\bibitem{Duplancic:2018bum}
G.~Duplančić, K.~Passek-Kumerički, B.~Pire, L.~Szymanowski, and S.~Wallon.
\newblock {Probing axial quark generalized parton distributions through
  exclusive photoproduction of a $\gamma\,\pi^\pm$ pair with a large invariant
  mass}.
\newblock {\em JHEP}, 11:179, 2018.

\bibitem{Grocholski:2021man}
Oskar Grocholski, Bernard Pire, Pawe\l{} Sznajder, Lech Szymanowski, and Jakub
  Wagner.
\newblock {Collinear factorization of diphoton photoproduction at next to
  leading order}.
\newblock {\em Phys. Rev. D}, 104(11):114006, 2021.

\bibitem{Qiu:2022bpq}
Jian-Wei Qiu and Zhite Yu.
\newblock {Exclusive production of a pair of high transverse momentum photons
  in pion-nucleon collisions for extracting generalized parton distributions}.
\newblock {\em JHEP}, 08:103, 2022.

\bibitem{Qiu:2022pla}
Jian-Wei Qiu and Zhite Yu.
\newblock {Single diffractive hard exclusive processes for the study of
  generalized parton distributions}.
\newblock {\em Phys. Rev. D}, 107(1):014007, 2023.

\bibitem{Qiu:2023mrm}
Jian-Wei Qiu and Zhite Yu.
\newblock {Extraction of the Parton Momentum-Fraction Dependence of Generalized
  Parton Distributions from Exclusive Photoproduction}.
\newblock {\em Phys. Rev. Lett.}, 131(16):161902, 2023.

\bibitem{Girod:2007aa}
F.X. Girod et~al.
\newblock {Measurement of Deeply virtual Compton scattering beam-spin
  asymmetries}.
\newblock {\em Phys.Rev.Lett.}, 100:162002, 2008.

\bibitem{HERMES:2012gbh}
A.~Airapetian et~al.
\newblock {Beam-helicity and beam-charge asymmetries associated with deeply
  virtual Compton scattering on the unpolarised proton}.
\newblock {\em JHEP}, 07:032, 2012.

\bibitem{Jo:2015ema}
H.~S. Jo et~al.
\newblock {Cross sections for the exclusive photon electroproduction on the
  proton and Generalized Parton Distributions}.
\newblock {\em Phys. Rev. Lett.}, 115(21):212003, 2015.

\bibitem{Defurne:2015kxq}
M.~Defurne et~al.
\newblock {E00-110 experiment at Jefferson Lab Hall A: Deeply virtual Compton
  scattering off the proton at 6 GeV}.
\newblock {\em Phys. Rev.}, C92(5):055202, 2015.

\bibitem{Defurne:2017paw}
M.~Defurne et~al.
\newblock {A glimpse of gluons through deeply virtual compton scattering on the
  proton}.
\newblock {\em Nature Commun.}, 8(1):1408, 2017.

\bibitem{COMPASS:2018pup}
R.~Akhunzyanov et~al.
\newblock {Transverse extension of partons in the proton probed in the
  sea-quark range by measuring the DVCS cross section}.
\newblock {\em Phys. Lett. B}, 793:188--194, 2019.
\newblock [Erratum: Phys.Lett.B 800, 135129 (2020)].

\bibitem{JeffersonLabHallA:2022pnx}
F.~Georges et~al.
\newblock {Deeply Virtual Compton Scattering Cross Section at High Bjorken xB}.
\newblock {\em Phys. Rev. Lett.}, 128(25):252002, 2022.

\bibitem{CLAS:2022syx}
G.~Christiaens et~al.
\newblock {First CLAS12 Measurement of Deeply Virtual Compton Scattering
  Beam-Spin Asymmetries in the Extended Valence Region}.
\newblock {\em Phys. Rev. Lett.}, 130(21):211902, 2023.

\bibitem{Braun:2020yib}
V.~M. Braun, A.~N. Manashov, S.~Moch, and J.~Schoenleber.
\newblock {Two-loop coefficient function for DVCS: vector contributions}.
\newblock {\em JHEP}, 09:117, 2020.

\bibitem{Braun:2012hq}
V.M. Braun, A.N. Manashov, and B.~Pirnay.
\newblock {Finite-t and target mass corrections to deeply virtual Compton
  scattering}.
\newblock {\em Phys.Rev.Lett.}, 109:242001, 2012.

\bibitem{Bertone:2021yyz}
V.~Bertone, H.~Dutrieux, C.~Mezrag, H.~Moutarde, and P.~Sznajder.
\newblock {The deconvolution problem of deeply virtual Compton scattering}.
\newblock {\em Phys. Rev. D}, 103(11):114019, 4 2021.

\bibitem{Dutrieux:2021wll}
H.~Dutrieux, H.~Dutrieux, O.~Grocholski, O.~Grocholski, H.~Moutarde,
  H.~Moutarde, P.~Sznajder, and P.~Sznajder.
\newblock {Artificial neural network modelling of generalised parton
  distributions}.
\newblock {\em Eur. Phys. J. C}, 82(3):252, 2022.
\newblock [Erratum: Eur.Phys.J.C 82, 389 (2022)].

\bibitem{Anikin:2007yh}
I.~V. Anikin and O.~V. Teryaev.
\newblock {Dispersion relations and subtractions in hard exclusive processes}.
\newblock {\em Phys. Rev. D}, 76:056007, 2007.

\bibitem{Polyakov:1999gs}
Maxim~V. Polyakov and C.~Weiss.
\newblock {Skewed and double distributions in pion and nucleon}.
\newblock {\em Phys.Rev.}, D60:114017, 1999.

\bibitem{Bakker:2004ib}
B.~L.~G. Bakker, E.~Leader, and T.~L. Trueman.
\newblock {A Critique of the angular momentum sum rules and a new angular
  momentum sum rule}.
\newblock {\em Phys. Rev. D}, 70:114001, 2004.

\bibitem{Leader:2013jra}
E.~Leader and C.~Lorc\'e.
\newblock {The angular momentum controversy: What\textquoteright{}s it all
  about and does it matter?}
\newblock {\em Phys. Rept.}, 541(3):163--248, 2014.

\bibitem{Polyakov:2002yz}
M.~V. Polyakov.
\newblock {Generalized parton distributions and strong forces inside nucleons
  and nuclei}.
\newblock {\em Phys. Lett.}, B555:57--62, 2003.

\bibitem{Diehl:2003ny}
M.~Diehl.
\newblock {Generalized parton distributions}.
\newblock {\em Phys.Rept.}, 388:41--277, 2003.

\bibitem{Ji:1998pc}
Xiang-Dong Ji.
\newblock {Off forward parton distributions}.
\newblock {\em J.Phys.}, G24:1181--1205, 1998.

\bibitem{Radyushkin:1998bz}
A.V. Radyushkin.
\newblock {Symmetries and structure of skewed and double distributions}.
\newblock {\em Phys.Lett.}, B449:81--88, 1999.

\bibitem{Chouika:2017dhe}
N.~Chouika, C.~Mezrag, H.~Moutarde, and J.~Rodríguez-Quintero.
\newblock {Covariant Extension of the GPD overlap representation at low Fock
  states}.
\newblock {\em Eur. Phys. J.}, C77:906, 2017.

\bibitem{Mezrag:2022pqk}
C\'edric Mezrag.
\newblock {An Introductory Lecture on Generalised Parton Distributions}.
\newblock {\em Few Body Syst.}, 63(3):62, 2022.

\bibitem{Chavez:2021llq}
Jos\'e Manuel~Morgado Chavez, Valerio Bertone, Feliciano De~Soto~Borrero,
  Maxime Defurne, C\'edric Mezrag, Herv\'e Moutarde, Jos\'e
  Rodr\'\i{}guez-Quintero, and Jorge Segovia.
\newblock {Pion generalized parton distributions: A path toward phenomenology}.
\newblock {\em Phys. Rev. D}, 105(9):094012, 2022.

\bibitem{Teryaev:2005uj}
O.~V. Teryaev.
\newblock {Analytic properties of hard exclusive amplitudes}.
\newblock In {\em {11th International Conference on Elastic and Diffractive
  Scattering: Towards High Energy Frontiers: The 20th Anniversary of the Blois
  Workshops, 17th Rencontre de Blois}}, 10 2005.

\bibitem{Diehl:2007jb}
M.~Diehl and D.~Yu. Ivanov.
\newblock {Dispersion representations for hard exclusive processes: beyond the
  Born approximation}.
\newblock {\em Eur. Phys. J. C}, 52:919--932, 2007.

\bibitem{Moutarde:2019tqa}
H.~Moutarde, P.~Sznajder, and J.~Wagner.
\newblock {Unbiased determination of DVCS Compton Form Factors}.
\newblock {\em Eur. Phys. J.}, C79(7):614, 2019.

\bibitem{Accardi:2012qut}
A.~Accardi et~al.
\newblock {Electron Ion Collider: The Next QCD Frontier}.
\newblock {\em Eur. Phys. J.}, A52(9):268, 2016.

\bibitem{AbdulKhalek:2021gbh}
R.~Abdul~Khalek et~al.
\newblock {Science Requirements and Detector Concepts for the Electron-Ion
  Collider: EIC Yellow Report}.
\newblock {\em arxiv:2103.05419}, 3 2021.

\bibitem{Chen:2018wyz}
Xurong Chen.
\newblock {A Plan for Electron Ion Collider in China}.
\newblock {\em PoS}, DIS2018:170, 2018.

\bibitem{Anderle:2021wcy}
Daniele~P. Anderle et~al.
\newblock {Electron-ion collider in China}.
\newblock {\em Front. Phys. (Beijing)}, 16(6):64701, 2021.

\bibitem{LHeCStudyGroup:2012zhm}
J.~L. Abelleira~Fernandez et~al.
\newblock {A Large Hadron Electron Collider at CERN: Report on the Physics and
  Design Concepts for Machine and Detector}.
\newblock {\em J. Phys. G}, 39:075001, 2012.

\bibitem{Pasquini:2014vua}
B.~Pasquini, M.~V. Polyakov, and M.~Vanderhaeghen.
\newblock {Dispersive evaluation of the D-term form factor in deeply virtual
  Compton scattering}.
\newblock {\em Phys. Lett. B}, 739:133--138, 2014.

\bibitem{Nussenzveig:1972tcd}
H.~M. Nussenzveig.
\newblock {\em {Causality and dispersion relations}}, volume~95.
\newblock Academic Press, New York, London, 1972.

\bibitem{Goldstein:2009ks}
Gary~R. Goldstein and Simonetta Liuti.
\newblock {The Use of Dispersion Relations in Hard Exclusive Processes and the
  Partonic Interpretation of Deeply Virtual Compton Scattering}.
\newblock {\em Phys. Rev. D}, 80:071501, 2009.

\bibitem{Cuic:2020iwt}
Marija \v{C}ui\'c, Kre\v{s}imir Kumeri\v{c}ki, and Andreas Sch\"afer.
\newblock {Separation of Quark Flavors Using Deeply Virtual Compton Scattering
  Data}.
\newblock {\em Phys. Rev. Lett.}, 125(23):232005, 2020.

\bibitem{Guidal:2002kt}
M.~Guidal and M.~Vanderhaeghen.
\newblock {Double deeply virtual Compton scattering off the nucleon}.
\newblock {\em Phys. Rev. Lett.}, 90:012001, 2003.

\bibitem{Belitsky:2003fj}
Andrei~V. Belitsky and Dieter Mueller.
\newblock {Probing generalized parton distributions with electroproduction of
  lepton pairs off the nucleon}.
\newblock {\em Phys. Rev. D}, 68:116005, 2003.

\bibitem{Deja:2023ahc}
K.~Deja, V.~Martinez-Fernandez, B.~Pire, P.~Sznajder, and J.~Wagner.
\newblock {Phenomenology of double deeply virtual Compton scattering in the era
  of new experiments}.
\newblock {\em Phys. Rev. D}, 107(9):094035, 2023.

\bibitem{Pedrak:2020mfm}
A.~Pedrak, B.~Pire, L.~Szymanowski, and J.~Wagner.
\newblock {Electroproduction of a large invariant mass photon pair}.
\newblock {\em Phys. Rev. D}, 101(11):114027, 2020.

\bibitem{Grocholski:2022rqj}
Oskar Grocholski, Bernard Pire, Pawe\l{} Sznajder, Lech Szymanowski, and Jakub
  Wagner.
\newblock {Phenomenology of diphoton photoproduction at next-to-leading order}.
\newblock {\em Phys. Rev. D}, 105(9):094025, 2022.

\bibitem{Duplancic:2023kwe}
Goran Duplan\v{c}i\'c, Saad Nabeebaccus, Kornelija Passek-Kumeri\v{c}ki,
  Bernard Pire, Lech Szymanowski, and Samuel Wallon.
\newblock {Probing chiral-even and chiral-odd leading twist quark generalized
  parton distributions through the exclusive photoproduction of a
  \ensuremath{\gamma}\ensuremath{\rho} pair}.
\newblock {\em Phys. Rev. D}, 107(9):094023, 2023.

\bibitem{DallOlio:2024vjv}
P.~Dall'Olio, F.~De~Soto, C.~Mezrag, J.~M. Morgado~Ch\'avez, H.~Moutarde,
  J.~Rodr\'\i{}guez-Quintero, P.~Sznajder, and J.~Segovia.
\newblock {Unraveling generalized parton distributions through Lorentz symmetry
  and partial DGLAP knowledge}.
\newblock {\em Phys. Rev. D}, 109(9):096013, 2024.

\bibitem{HadStruc:2024rix}
Herv\'e Dutrieux, Robert~G. Edwards, Colin Egerer, Joseph Karpie, Christopher
  Monahan, Kostas Orginos, Anatoly Radyushkin, David Richards, Eloy Romero, and
  Savvas Zafeiropoulos.
\newblock {Towards unpolarized GPDs from pseudo-distributions}.
\newblock {\em JHEP}, 08:162, 2024.

\bibitem{Accardi:2020swt}
A.~Accardi et~al.
\newblock {An experimental program with high duty-cycle polarized and
  unpolarized positron beams at Jefferson Lab}.
\newblock {\em Eur. Phys. J. A}, 57(8):261, 2021.

\bibitem{Dutrieux:2021ehx}
H.~Dutrieux, V.~Bertone, H.~Moutarde, and P.~Sznajder.
\newblock {Impact of a positron beam at JLab on an unbiased determination of
  DVCS Compton form factors}.
\newblock {\em Eur. Phys. J. A}, 57(8):250, 2021.

\bibitem{GK}
R.~Gnanadesikan and J.~R. Kettenring.
\newblock Robust estimates, residuals, and outlier detection with multiresponse
  data.
\newblock {\em Biometrics}, 28(1):81--124, 1972.

\bibitem{Rousseeuw}
Peter~J. Rousseeuw.
\newblock Least median of squares regression.
\newblock {\em Journal of the American Statistical Association},
  79(388):871--880, 1984.

\bibitem{LEDOIT}
Olivier Ledoit and Michael Wolf.
\newblock A well-conditioned estimator for large-dimensional covariance
  matrices.
\newblock {\em Journal of Multivariate Analysis}, 88(2):365--411, 2004.

\bibitem{Dutfuture}
Herv\'e Dutrieux.
\newblock Robust covariance and extremely correlated fit.
\newblock {\em in preparation}, 2024.

\bibitem{Bertone:2013vaa}
V.~Bertone, S.~Carrazza, and J.~Rojo.
\newblock {APFEL: A PDF Evolution Library with QED corrections}.
\newblock {\em Comput. Phys. Commun.}, 185:1647--1668, 2014.

\bibitem{Bertone:2017gds}
Valerio Bertone.
\newblock {APFEL++: A new PDF evolution library in C++}.
\newblock {\em PoS}, DIS2017:201, 2018.

\bibitem{Berthou:2015oaw}
B.~Berthou, D.~Binosi, N.~Chouika, L.~Colaneri, M.~Guidal, C.~Mezrag,
  H.~Moutarde, J.~Rodríguez-Quintero, F.~Sabatié, P.~Sznajder, and J.~Wagner.
\newblock {PARTONS: PARtonic Tomography Of Nucleon Software. A computing
  framework for the phenomenology of Generalized Parton Distributions}.
\newblock {\em Eur. Phys. J.}, C78(6):478, 2018.

\end{thebibliography}
